\documentclass[reprint, amsmath,amssymb, aps,pra,superscriptaddress]{revtex4-2}

\usepackage{graphicx}
\usepackage{dcolumn}
\usepackage{bm}
\usepackage[margin=2.0cm]{geometry}
\usepackage{amsmath}
\usepackage{braket}
\usepackage{appendix}
\usepackage{float}
\usepackage{siunitx}
\usepackage{amssymb}
\usepackage{color}
\usepackage{multirow}
\usepackage{longtable}
\usepackage{ulem}
\usepackage{soul}

\usepackage{silence}
\begin{document}

\title{ Laser-target symmetry-breaking in high harmonic generation: from frequency shift to odd-even intensity modulation 
}
\author{Doan-An Trieu}
    \affiliation{Computational Physics Key Laboratory K002, Department of Physics, Ho Chi Minh City University of Education, Ho Chi Minh City 72711, Vietnam}
    
\author{Van-Hoang Le}
    \affiliation{Computational Physics Key Laboratory K002, Department of Physics, Ho Chi Minh City University of Education, Ho Chi Minh City 72711, Vietnam} 
\author{Ngoc-Loan Phan}
    \email{loanptn@hcmue.edu.vn}
    \affiliation{Computational Physics Key Laboratory K002, Department of Physics, Ho Chi Minh City University of Education, Ho Chi Minh City 72711, Vietnam}

\date{\today}

\begin{abstract}

Although the frequency shift and odd-even intensity modulation in high-order harmonic generation (HHG) have both been observed for asymmetric laser-target systems, they are typically studied as two separate phenomena. In this Letter, we provide a comprehensive picture of these two nonlinear optical phenomena, unifying them through a common origin - asymmetry of the laser-target system. By tuning asymmetric laser-target systems, we discover a transition from the harmonic frequency shift to the odd-even intensity modulation upon increasing the duration of the driving laser pulse. Specifically, these phenomena are observed simultaneously for laser pulses with intermediate pulse duration. For numerical evidence, we solve the time-dependent Schr\"{o}dinger equation, while insight into the underlying physics is obtained from a simplified analytically tractable model. Understanding the asymmetric characteristics reflected in the HHG as provided is crucial for retrieving laser-target information, sampling external fields, and probing molecular dynamics.

\end{abstract}

\maketitle

\textit{Introduction ---}  High-order harmonic generation (HHG) resulting from the laser-matter interaction is crucial for attosecond technologies, enabling tracking ultrafast dynamics inside matters with unprecedented temporal resolution~\cite{Brabec:RMP00,Corkum:NatPhys17,Ramasesha:ARPC16,Nisoli:ChemRev17,Li:NatCom20}. On the other hand, the HHG by itself reflects various properties of the laser-target system; therefore, it is an effective tool for imaging molecular structures and dynamics~\cite{Itatani:Nat04,Haessler:NatPhys10,Lan:prl17,Kim:NatPho14,Peng:NatRevPhys:19}. Fully comprehending HHG features benefits these important applications in strong-field physics and attosecond sciences.

A well-known feature of HHG emitted from a symmetric laser-target system is its spectrum peaks positioning at only odd multiples of the fundamental laser frequency \cite{Corkum:prl93,Lewenstein:pra94}. This pattern results from the interference of uniform attosecond bursts emitted every half-cycle time intervals as the electron recombines with the parent ion after being driven by the laser electric field following the tunneling ionization~\cite{Corkum:prl93,Lewenstein:pra94,Mairesse:Sc03,Niikura:prl03}. Meanwhile, for a laser-target system lacking centrosymmetric symmetry, HHG reflects this symmetry-breaking by producing even-order harmonics in addition to the odd-order ones~\cite{Ben-Tal_1993,Neufeld:NatCom19,Frumker:PRL12,Kraus:prl12,Chen:prl13,Nguyen:ComPhy20,Phan:PCCP19,Nguyen:pra22,Ngan:pra20,Luu:NatCom18,Uzan:prl23,Li:NC23,Lambert:NatCom15,Vampa:NP18,Luu:pra18,Shafir:NatPhys09,Niikura:prl10,Li:pra20,Trieu:PRA23}. Tuning asymmetry of a laser-target system thus modulates the intensities of odd and even harmonics~\cite{Chen:prl13,Nguyen:ComPhy20,Phan:PCCP19,Nguyen:pra22,Ngan:pra20,Luu:NatCom18,Uzan:prl23,Li:NC23,Lambert:NatCom15,Vampa:NP18,Luu:pra18,Shafir:NatPhys09,Niikura:prl10,Li:pra20,Trieu:PRA23}.

Recent advances in laser technology have progressively shortened laser pulses to just a few cycles with stable carrier-envelope phases (CEP)~\cite{Paulus:prl03,Wittmann:NatPhys09,Hauri:aplB04}. Interacting with those pulses, targets emit HHG, which no longer follows the odd- or even-order rules of harmonics due to the nonuniform attosecond bursts caused by the rapid change of the instantaneous electric field after half of the laser cycle~\cite{Lan:prl17,Naumov:pra15}. Furthermore, adjusting the short-laser-pulse's asymmetry shifts the harmonic peaks, which is significant for characterizing laser CEP~\cite{Naumov:pra15,Hollinger:oe20} and pulse waveform~\cite{You:OL17,Trieu:24}, probing nuclear dynamics~\cite{Lan:prl17,Bian:prl14}, and controlling the ionization and recombination times~\cite{Shafir:NatPhys09,Kim:NatPho14}.    

While the odd-even intensity modulation~\cite{Frumker:PRL12, Kraus:prl12,Chen:prl13,Nguyen:ComPhy20, Phan:PCCP19,Nguyen:pra22,Ngan:pra20,Luu:NatCom18,Uzan:prl23,Li:NC23,Lambert:NatCom15,Vampa:NP18,Luu:pra18,Shafir:NatPhys09,Niikura:prl10,Li:pra20,Trieu:PRA23} and harmonic frequency shift~\cite{Zhou:prl96,Graml:PRB23, Haworth:NatPhys07,Naumov:pra15,You:OL17,Schmid:Nat21,Schubert:NP14,Li:pra20,Mandal:OE22,Hollinger:oe20,Lan:prl17,Bian:prl14,Shafir:NatPhys09,Kim:NatPho14,Shafir:Nat12,Silaev:JPCS22,Trieu:24} both characterize asymmetric laser-target systems in two limits of the laser pulse, multi-cycle versus few-cycle, they are usually studied separately since different laser-target objects are considered. Remarkably, there is still no direct relationship between the rules governing these two phenomena. Also, it is intriguing to identify situations where both effects are simultaneously observable.

In the present Letter, we theoretically study the link between harmonic frequency shift and odd-even intensity modulation upon increasing the duration of the laser pulses while interacting with tunable asymmetric laser-target systems. The two systems with different controlling symmetry-breaking factors are considered: (i) a symmetric laser-atom system with an additional low-frequency electric field, and (ii) a planar molecule with varying orientation angles. We demonstrate that harmonic frequency shift and odd-even harmonic modulation are both governed by the same symmetry-breaking factor and, indeed, are the two different manifestations of this symmetry-breaking in HHG. Also, we observe both effects occurring simultaneously in HHG at intermediate pulse durations.

\begin{figure*} [tb!]
    \centering
    \includegraphics[width=0.6\linewidth]{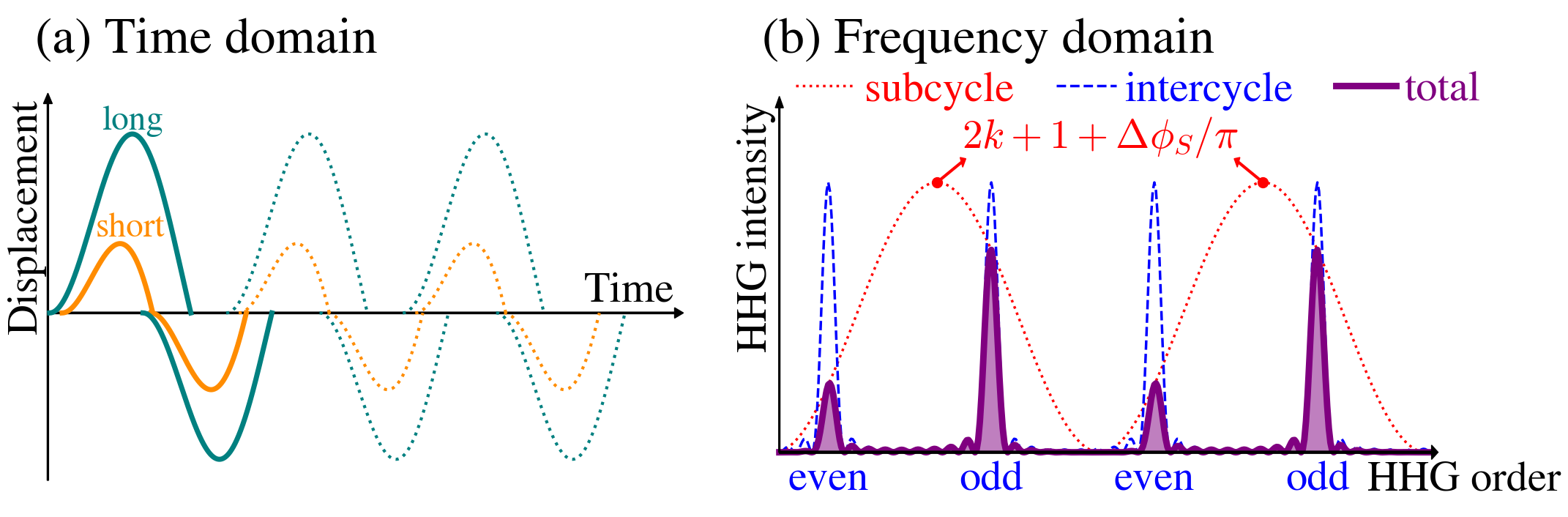}
    \caption{Sketch of subcycle and intercycle interference in the time~(a) and frequency~(b) domains. The subcycle interference is synthesized from subcycle emission bursts, emitted as electron recombination after implementing closed trajectories~[solid curves in Fig~(a)], resulting in an interference pattern with flat peaks whose positions may not at integer harmonic orders [red dotted curve in Fig.~(b)]. The intercycle interference resulting from emission bursts repeated each cycle (a) leads to a comb pattern [blue dashed curves in Fig~(b)]. In a multicycle laser pulse, both subcycle and intercycle are involved, leading to odd-even intensity modulation~[purple solid curves in Fig~(b)].   
    }
    \label{fig:mechanism}
\end{figure*}

\textit{Subcycle and intercycle interferences ---} 
Although harmonic frequency shift and odd-even intensity modulation have been studied separately before (see Refs.~\cite{Naumov:pra15,Ben-Tal_1993,Neufeld:NatCom19,Frumker:PRL12,You:OL17}), we broadly review them here to emphasize their interconnectedness and explain their underlying physics comprehensively. 

According to the strong-field approximation, HHG in the time domain is indeed the interference of attosecond bursts emitted at each recombination event of electrons to the parent ion after traveling in the laser electric field following the ionization near the laser intensity peaks~\cite{Corkum:prl93,Lewenstein:pra94}. As shown in Fig.~\ref{fig:mechanism}, two types of interference exist for asymmetric laser-target systems: subcycle and intercycle. \textit{Subcycle interference} arises from adjacent bursts emitted within one optical cycle. During each half-cycle, two bursts are generated by short and long trajectories responsible for generating each harmonics~\cite{Corkum:prl93,Lewenstein:pra94}. However, for harmonics near the cutoff or under good phase-matching, only short trajectories remain, while long trajectories cancel out each other. Thereby, the subcycle interference can be mathematically expressed as 
\begin{equation} \label{eq:D-sub}
 D_\mathrm{sub} (N)  = D_\mathrm{_1} e^{i \varphi_\mathrm{_1}} \bigg [1 - \dfrac{D_\mathrm{_2}}{D_\mathrm{_1}} e^{i(\Delta \varphi - N \pi)} \bigg],
\end{equation}
where $N$ is the harmonic order; $D_\mathrm{j}$ and $\varphi_\mathrm{j}$ are the amplitude and phase of the emission bursts; and the index $\mathrm{j} = 1,~2$ means the first and second half-cycles.
$\Delta \varphi = \varphi_\mathrm{2} - \varphi_\mathrm{1}$ is the phase difference between the two attosecond bursts. For symmetry-breaking laser-target systems, the two adjacent bursts with a half-cycle time interval are asymmetric both in amplitude and phase. 

Subcycle interference [Eq.~(\ref{eq:D-sub})] modulates its intensity as presented by the red dotted curve in Fig.~\ref{fig:mechanism}(b), whose maxima locate at harmonic orders
\begin{equation} \label{eq:shift}
    N_\mathrm{sub-max} =  2m+1 + \dfrac{\Delta \varphi}{\pi},
\end{equation}
which strongly depend on the phase difference $\Delta \varphi$, characterizing the asymmetry of the laser-target system. Here, $m$ is an integer. If the system is centrosymmetric under the half-cycle time translation ($\Delta \varphi=0$), the maxima overlap with odd-order harmonics only. 

Subcycle interference occurs when the laser pulse consists of a few optical cycles, i.e., in the few-cycle limit. However, for a multicycle laser pulse, the subcycle emission bursts are repeated every cycle but weighted by the factor $\alpha$ because of the unequal instantaneous ionization rate, unequal peak amplitudes of electric fields in each optical cycle, or the depletion.
In this case, the intercycle interference between subcycle emissions bursts is involved as 
\begin{equation} \label{eq:D-multi}
    D(N) = D_\mathrm{sub} (N) ~ D_\mathrm{inter} (N).
\end{equation}
where $ D_\mathrm{inter} (N) = \sum\limits_{j=0}^{n} \alpha_j e^{j i N \omega T}$
characterizes the \textit{intercycle interference}. Here, $n$ is the number of the pulse's optical cycles, $\omega$ and $T$ are the laser carrier frequency and optical cycle. In the continuous-wave limit or the number cycle is large enough, every optical cycle is approximately identical, i.e., $\alpha_j \approx 1$, the intercycle interference has magnitude  
\begin{equation} \label{eq:D_int}
  |D_\mathrm{inter-cw} (N)|    = \dfrac{\sin [(n+1) N \pi]}{\sin ( N \pi)}.
\end{equation}
Equation~(\ref{eq:D_int}) implies that the peaks are placed at integer numbers of $N$, i.e., at both odd- and even-order harmonics, as presented by the dashed blue curve in Fig.~\ref{fig:mechanism}(b).

Equation~(\ref{eq:D-multi}) demonstrates that when an asymmetric laser-target system interacts with a multicycle laser pulse, the harmonic peak's frequency (odd and even) is governed by $D_\mathrm{inter} (N)$, but amplitude is controlled by $D_\mathrm{sub} (N)$. Thereby, the intensity ratio between the even and odd harmonics is 
\begin{equation} \label{eq:e2o}
    \dfrac{I_\mathrm{even}}{I_\mathrm{odd}} \approx \dfrac{|D_\mathrm{sub} (2m)|^2} {|D_\mathrm{sub} (2m+1)|^2} \approx 1 - \dfrac{2 \kappa \cos \Delta \varphi } {1+\kappa \cos\Delta \varphi },
\end{equation}
where $\kappa = \dfrac{2 D_\mathrm{1} D_\mathrm{2}}{|D_\mathrm{1}|^2 + |D_\mathrm{2}|^2}$ is the intensity imbalance of the two subcycle emission bursts. This odd-even intensity was well validated for polar molecules in a multicycle laser pulse~\cite{Ngan:pra20}, or atoms in a combination of a multicycle pulse with a low-frequency electric pulse~\cite{Trieu:PRA23}. 

In summary, we have connected two nonlinear optical phenomena, the harmonic frequency shift and the odd-even intensity modulation, as manifestations of the phase difference between adjacent attosecond bursts. Their appearance depends on the duration of the laser pulse. In a few-cycle laser pulse, only subcycle interference is involved in the HHG of an asymmetric laser-target system. The symmetry-breaking factor causes a harmonic frequency shift by Eq.~\eqref{eq:shift}. In contrast, in multicycle laser pulse, both subcycle and intercycle interferences are involved, producing both odd and even harmonics, whose even-to-odd ratio is modified by changing the asymmetry factor by Eq.~\eqref{eq:e2o}. The governing factor of these two effects is the same: the phase difference between adjacent attosecond bursts, $\Delta \varphi$. 

\begin{figure*} [tb!]
    \centering
    \includegraphics[width=0.9\linewidth]{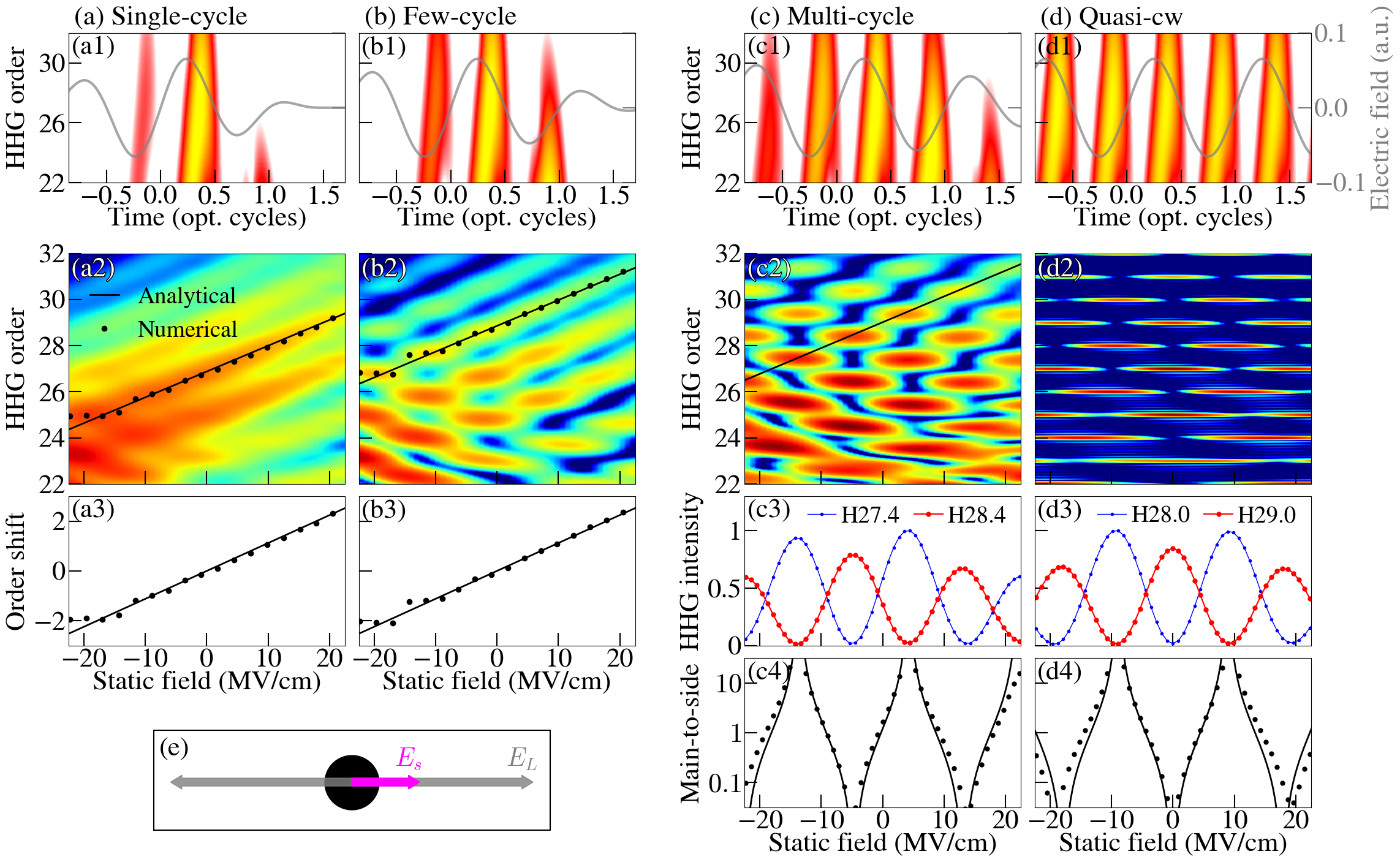}
    \caption{Transition from SEF-induced harmonic frequency shift in single-cycle (a) and few-cycle (b) regimes to SEF-induced odd-even intensity modulation in multi-cycle (c) and quasi-cw (d) regimes of HHG emitted from a hydrogen atom in the combination of driving laser pulse and SEF~(e). Upper panels [(a1) - (d1)] show the electric field of driving laser pulses (grey curves) and the harmonic emission bursts in the absence of  SEF.
    Panels [(a2) - (d2)] demonstrate the TDSE-simulated HHG for harmonics near cutoff when varying SEF.
    In Figs. [(a2), (b2)], the black dots pick the centroids of harmonic peaks, which closely follow the analytical predictions~(\ref{eq:THz-shift}) presented by solid lines. The separated comparisons are exhibited in Figs.~[(a3), (b3)]. For multi-cycle and quasi-cw regimes, adding a SEF causes additional harmonics (called sideband) besides existing peaks without SEF (called mainband) [(c2), (d2)]. Varying SEF, the intensity of the mainband and the sideband [(c3), (d3)] and the even-to-odd ratios [(c4), (d4)] modulate with the same period, which is consistent with the analytical prediction~(\ref{eq:e2o-THz}) (black curves). When using intermediate pulses [(b), (c)], both harmonic frequency shift and odd-even intensity modulation are observed.
    The driving laser with the intensity of $1.5\times 10^{14}$ W/cm$^2$, and the wavelength of 800~nm is used. }
    \label{fig:Overviews}
\end{figure*}

\textit{Transition from harmonic frequency shift to odd-even intensity modulation ---} To investigate the interconnectedness between the harmonic frequency shift and odd-even intensity modulation in HHG, we consider laser-target systems with varying laser pulse durations and tunable symmetry-breaking factors. Specifically, we show results for driving laser pulses with three, four, and six optical cycles, called single-cycle, few-cycle, and multi-cycle regimes. In addition, the quasi-continuous-wave limit (quasi-cw) is also given using a driving laser pulse of ten optical cycles with the flattop envelope. The cosine-squared envelope and carrier-envelope phase (CEP) of $-\pi/2$ are chosen for all the considered driving pulses. Two targets are considered: (i) a hydrogen atom with an additional weak low-frequency electric field and (ii) a planar molecule H$_3^{2+}$ with varying orientation angles. We chose these cases because the symmetry-breaking factors (the external electric field and the molecular orientation) causing the asymmetry are tunable. In the adiabatic approximation, a low-frequency electric field (terahertz field) can be considered as a quasi-static electric field.

We proceed the study with numerical evidence from solving the time-dependent Schr\"{o}dinger equation (TDSE). These results can be additionally validated by the analytical analysis based on simple models. We show the results in the main text while the details of calculations are provided in supplementary~\cite{suppl}.  

\vspace{0.15cm}
\noindent \textit{(i) Quasi-static electric field as a symmetry-breaking factor}

\vspace{0.1 cm}

Figure~\ref{fig:Overviews} exhibits a scenario of HHG where a hydrogen atom interacts with the driving laser pulse combined with a tunable quasi-static electric field (SEF). Following the pulse-duration change from single-cycle to quasi-cw regimes, this figure demonstrates the transition from SEF-induced harmonic frequency shift [Panels (a)-(c)] to  SEF-induced even-to-odd modulation in the HHG [Panels (c)-(d)].

Particularly, in single-cycle and few-cycle regimes, there are only two or three attosecond bursts, as shown in the time domain [Panels (a1)-(b1)], consequently only subcycle interference occurs. In this case, harmonic peaks shift because of the varying SEF-induced asymmetry [Panels (a2)-(b2)]. The figures show that the shift is up to four harmonic orders when SEF varies from $-20$ to $+20$~MV/cm. Meanwhile, when the driving laser pulse is long enough (multi-cycle or quasi-cw), both subcycle and intercycle interference are involved because of the contribution of more emission bursts in the time domain [Panels (c1)-(d1)]. The intercycle interference sharpens the harmonic peaks at specific orders for the multi-cycle regimes or integers for the quasi-cw regime. Moreover, varying SEF leads to the intensity oscillation of these harmonic peaks, as shown in Panels (c3)-(d3), resulting in the oscillation of the intensity ratio between the adjacent harmonic orders, called even-to-odd ratio [Panels (c4)-(d4)].

Specifically, both phenomena are observed in the same HHG pattern in Panels (b) and (c), where the pulses with intermediate duration are used. The frequency shift is seen clearly for harmonics near and at the cutoff, while the odd-even intensity oscillation is found for harmonics below cutoffs. This simultaneous existence of the two effects can be explained by looking into the time-frequency profiles in Panels (b1)-(c1). They show that the harmonics near the cutoff coherently interfere from only two emission bursts, i.e., subcycle interference, leading to the harmonic frequency shift. Meanwhile, the low harmonics away from the cutoff result from more than two emission bursts, i.e., both subcycle and multi-cycle interference, causing odd-even intensity oscillation.  

To continue, we analyze the observed transition between the two effects,  from the harmonic frequency shift to the even-to-odd ratio oscillation. It is well-known that the asymmetry caused by the SEF leads to the phase distortion of the attosecond bursts. This SEF-induced phase distortion difference has been calculated analytically in Refs.~\cite{Trieu:PRA23,Trieu:24} within the strong-field approximation ~\cite{Corkum:prl93,Lewenstein:pra94}. Substituting it into Eq.~(\ref{eq:shift}), we can describe SEF-induced shift in the form 
\begin{equation} \label{eq:THz-shift}
    \Delta N = \pm  2 C \dfrac{E_m}{\pi \omega^3} E_T,
\end{equation}
where $ \Delta N$ is the shift of harmonic order; $E_m$ is the average amplitude of the two electric peaks at the two half-cycles at the center of the driving pulses, which are responsible for the emission bursts; $E_T$ is the SEF; and $C$ is a constant, which is equal to $2.558$ for harmonics near cutoff. 
The analytical prediction (\ref{eq:THz-shift}) of SEF-induced frequency shift is fairly consistent with that from numerical simulation, as shown in Panels (a2), (a3), (b2), (b3), and (c2). 

For the quasi-cw regime, the simulated even-to-odd ratio matches the analytical description well. This formula, which has the form 
\begin{equation} \label{eq:e2o-THz} 
    \dfrac{I_\mathrm{even}}{I_\mathrm{odd}} \approx \tan^2 \bigg(C \dfrac{E_0}{ \omega^3} 
 E_T \bigg),
\end{equation}
is a consequence of Eq.~(\ref{eq:e2o}) incorporated with the analytical THz-induced phase distortion difference. The demonstration of Eq.~\eqref{eq:e2o-THz} in Panels (c4) and (d4) shows its consistency with numerical results. It is worth noting that the even-to-odd ratio (\ref{eq:e2o-THz}) and the frequency shift ~(\ref{eq:THz-shift}) are the two manifestations of the same symmetry-breaking nature. The common factor $2C E_0 / \pi \omega^3$ is simultaneously the slope of linear frequency shift in the single- or few-cycle regimes and half of the oscillation frequency in odd-even intensity oscillation in the multi-cycle and quasi-cw regimes. 

\vspace{0.15cm}
\noindent \textit{(ii) Orientation of planar molecule H$_3^{2+}$ as a symmetry-breaking factor}

\vspace{0.1 cm}

\begin{figure*}  [tb!]
    \centering
    \includegraphics[width=0.85\linewidth]{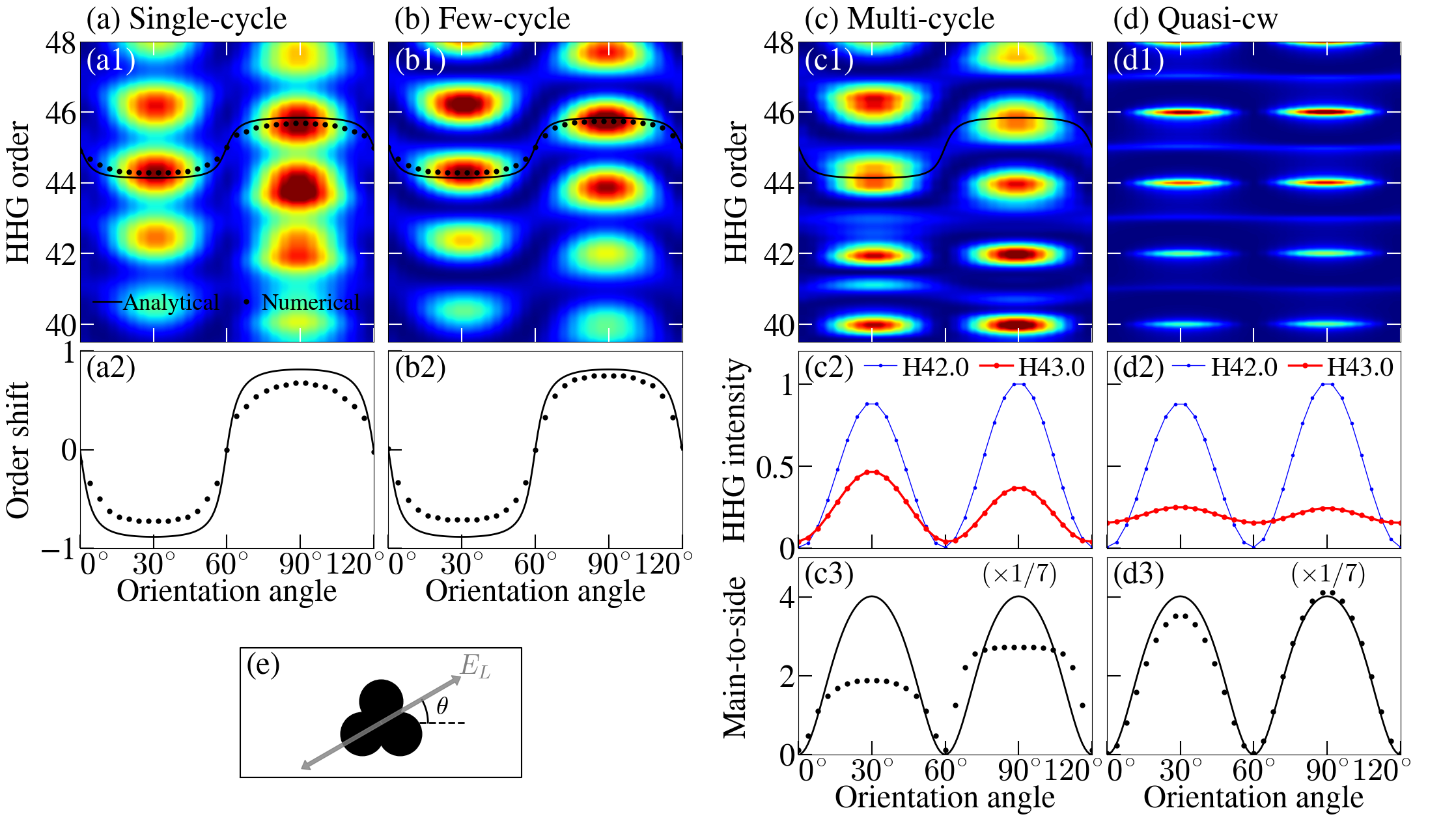}
    \caption{Transition from harmonic frequency shift~[(a)-(c)] to odd-even intensity oscillation~[(c)-(d)] for a planar molecule H$_3^{2+}$ with varying orientation angle [model in Fig.~(e)], interacting with a linearly polarized laser pulse in single-, few-, multi-cycle, and quasi-cw regimes. When using a multi-cycle laser (c), both frequency shift and odd-even oscillation are simultaneously appeared. In Figs.~[(a2), (b2), (c3), (d3)], the frequency shift and odd-even oscillation numerically simulated from TDSE (black dots) fairly match with those predicted analytically (solid curves). The laser with the intensity of $3\times 10^{14}$ W/cm$^2$, and the wavelength of 800~nm is used.
    }
    \label{fig:H3}
\end{figure*}

Figure~\ref{fig:H3} exhibits the numerical simulation of HHG obtained by solving TDSE for planar molecule H$_3^{2+}$ with varying orientation angles in different pulse regimes. Because of the C$_3$ symmetry of H$_3^{2+}$, we present the results only from $0^\circ$ to $120^\circ$ for the orientation angle. The first observation from this picture is that, for all regimes, the HHG intensity is enhanced at orientation angles of $30^\circ$ and $90^\circ$, and strongly suppressed at $0^\circ$, $60^\circ$, and $120^\circ$. This observation is similar to those observed in C$_3$-symmetry monocrystal~\cite{Kong:FIP22,Omasaka:oc22}. The HHG orientation dependence is easily understood by its consistency with the orientation-dependent electron density, as seen in Panel~(e). 

The second feature is the transition from the nonlinear frequency shift [Panels (a)-(c)] into the odd-even intensity modulation [Panels (c)-(d)] with varying orientation while increasing the pulse duration. Particularly, as a result of the subcycle interference, the harmonic frequencies nonlinearly shift with the period of $120^\circ$ in the near-cycle and few-cycle regimes [Panels (a)-(b)]. To the best of our knowledge, most harmonic frequency shifts observed so far vary linearly with the changing of asymmetry factors~\cite{Trieu:24,Naumov:pra15,You:OL17,Schmid:Nat21,Schubert:NP14,Hollinger:oe20,Mandal:OE22}. This nonlinear harmonic frequency shift is intriguing. On the other hand, for harmonics in the quasi-cw regime [Panels (c)-(d)], the positions of harmonic peaks are fixed, but the harmonic intensity is modulated with the period of $60^\circ$ when varying molecular orientation. This odd-even intensity modulation is a consequence of both subcycle and intercycle interferences.

With a laser pulse of intermediate duration, i.e., in the multi-cycle regime as shown in Panels (c), both frequency shift and odd-even modulation are observed. Similar to the case of SEF-induced symmetry-breaking, the harmonic frequency shift is clearly seen for harmonics at the cutoff, while low harmonics exhibit odd-even intensity modulation.
  
Our previously mentioned observations can be additionally justified by analytical means. According to Eqs.~(\ref{eq:shift}) and (\ref{eq:e2o}), for formulating analytical formulae of the frequency shift and even-to-odd ratio, we need an analytical form of the phase difference between the adjacent attosecond bursts $\Delta \varphi$. However, different from the SEF case where $\Delta \varphi$ results from the SEF-induced distortion of electron trajectories that can easily be treated by strong-field approximation, in the oriented H$_3^{2+}$-molecule case, $\Delta \varphi$ is caused by the molecule itself. Therefore, we adopt the method of linear combination of atomic orbitals (LCAO) to approximately describe the molecule H$_3^{2+}$. The details are presented in Supplementary~\cite{suppl}. 

According to LCAO, the electron recombination to the ground state gives an emission characterized as $D_1 \sim d_{\mathrm{1}} e^{i \varphi_{\mathrm{1}}}$, where the phase $\varphi_{\mathrm{1}}$ satisfies 
\begin{equation} \label{eq:phi_H3}
\tan \varphi_{\mathrm{1}} = \dfrac{\sin \mathbf{k}\mathbf{R}_1 + \sin\mathbf{k}\mathbf{R}_2 + \sin\mathbf{k}\mathbf{R}_3}{\cos\mathbf{k}\mathbf{R}_1 + \cos\mathbf{k}\mathbf{R}_2 + \cos\mathbf{k}\mathbf{R}_3},
\end{equation}
with $\mathbf{R}_j$ ($j=1$, 2, 3) are coordinates of the H atoms with respect to the center of charge, and $\mathbf{k}$ is the electron wave number. After half of an optical cycle, the second emission is $D_2 \sim d_{\mathrm{2}} e^{i (\varphi_{\mathrm{2}} + N\pi)}$ with $\varphi_{\mathrm{2}} = - \varphi_{\mathrm{1}}$.

First, we discuss the quasi-cw regime. Substituting $\Delta \varphi = - 2\varphi_{\mathrm{1}}$ and assume $|d_{\mathrm{_1}}| \approx |d_{\mathrm{2}}|$ for simplicity, in to Eq.~(\ref{eq:e2o}), we have the even-to-odd ratio
\begin{equation} \label{eq:e2o_H3}
    \dfrac{I_{\mathrm{even}}}{I_{\mathrm{odd}}} = \bigg| \dfrac{\sin{(y \sin \theta)} + \sin (y \cos \theta_1 ) - \sin(y \cos \theta_2 ) } {\cos{(y \sin \theta)}  + \cos (y \cos \theta_1) + \cos (y \cos \theta_2 )}  \bigg|^2,
\end{equation}
where $\theta$ is the orientation angle, $\theta_{1(2)} = \theta \pm \pi/6$, and $y = k |\mathbf{R}_j|$ with $k = \sqrt{2N \omega}$. Its plot in Panels (c3)-(d3) qualitatively follows the oscillation of even-to-odd ratio numerically simulated from TDSE. The scaling factor between analytical and numerical odd-even ratio is caused by the approximation $|d_{\mathrm{_1}}| \approx |d_{\mathrm{2}}|$, i.e., ignoring the magnitude asymmetry of the two adjacent attosecond bursts.

To analytically describe the frequency shift, we can not apply Eq.~(\ref{eq:shift}) directly because the phase $\varphi_1$ also depends on the harmonic order $N$ via Eq.~\eqref{eq:phi_H3}. This circumstance makes Eq.~(\ref{eq:shift}) nontrivial. Therefore, we approach this problem another way, rewriting Eq.~\eqref{eq:D-sub} for the subcycle interference, and find the condition that maximizes
\begin{equation} \label{eq:IsubH3}
     I_\mathrm{sub}  =  |D_1 - D_2|^2.
\end{equation}
Detailed partition of Eq.~(\ref{eq:IsubH3}) is shown in Supplementary~\cite{suppl}. Its solution is found numerically and presented in Panels (a) and (b), which fairly follows the results from the TDSE simulation.

\textit{Conclusions and discussions---} Depending on the laser pulse duration, the subcycle only or both subcycle and intercycle interferences of attosecond bursts occur, which is reflected in the HHG spectra via the harmonic frequency shift or even-to-odd intensity modulation. These non-linear effects are the two manifestations of one common physics - the asymmetry of laser-target systems. Although the harmonic shift and the even-to-odd harmonic intensity modulation have been investigated previously, for the first time we provide a unified picture for both of them, where the transition is demonstrated by varying the laser pulse duration from single to multi-cycles. Interestingly, both effects in HHG spectra are observable for the laser pulse with intermediate duration.    

For the above conclusion, we have utilized the numerical method of solving the Schr\"{o}dinger equation combined with the theoretical analysis based on simple models. This study adopts two laser-target systems with tunable symmetry-breaking factors; however, the results are general, and the conclusion can be used for different asymmetric laser-target systems. 

We can also discuss the application prospective because the harmonic frequency shift and even-to-odd intensity modulation clearly encode the laser-target information. For the first case in this study, where the electric field acts as a symmetry-breaking factor, the asymmetry manifestations in HHG can be leveraged to sample the field's temporal profile or control the subcycle motion of electron wavepackets. Meanwhile, when asymmetry factors originate from the targets themselves, the harmonic frequency shift and even-to-odd harmonic modulation can serve as tools to retrieve molecular structures or to probe molecular dynamics.

\section*{Acknowledgements}
This work was funded by Vingroup and supported by Vingroup Innovation Foundation (VINIF) under project code VINIF.2021.DA00031. The calculations were executed by the high-performance computing cluster at Ho Chi Minh City University of Education, Vietnam.

\bibliographystyle{apsrev4-1}
\bibliography{refs.bib}

\begin{thebibliography}{52}%
\makeatletter
\providecommand \@ifxundefined [1]{%
 \@ifx{#1\undefined}
}%
\providecommand \@ifnum [1]{%
 \ifnum #1\expandafter \@firstoftwo
 \else \expandafter \@secondoftwo
 \fi
}%
\providecommand \@ifx [1]{%
 \ifx #1\expandafter \@firstoftwo
 \else \expandafter \@secondoftwo
 \fi
}%
\providecommand \natexlab [1]{#1}%
\providecommand \enquote  [1]{``#1''}%
\providecommand \bibnamefont  [1]{#1}%
\providecommand \bibfnamefont [1]{#1}%
\providecommand \citenamefont [1]{#1}%
\providecommand \href@noop [0]{\@secondoftwo}%
\providecommand \href [0]{\begingroup \@sanitize@url \@href}%
\providecommand \@href[1]{\@@startlink{#1}\@@href}%
\providecommand \@@href[1]{\endgroup#1\@@endlink}%
\providecommand \@sanitize@url [0]{\catcode `\\12\catcode `\$12\catcode `\&12\catcode `\#12\catcode `\^12\catcode `\_12\catcode `\%12\relax}%
\providecommand \@@startlink[1]{}%
\providecommand \@@endlink[0]{}%
\providecommand \url  [0]{\begingroup\@sanitize@url \@url }%
\providecommand \@url [1]{\endgroup\@href {#1}{\urlprefix }}%
\providecommand \urlprefix  [0]{URL }%
\providecommand \Eprint [0]{\href }%
\providecommand \doibase [0]{http://dx.doi.org/}%
\providecommand \selectlanguage [0]{\@gobble}%
\providecommand \bibinfo  [0]{\@secondoftwo}%
\providecommand \bibfield  [0]{\@secondoftwo}%
\providecommand \translation [1]{[#1]}%
\providecommand \BibitemOpen [0]{}%
\providecommand \bibitemStop [0]{}%
\providecommand \bibitemNoStop [0]{.\EOS\space}%
\providecommand \EOS [0]{\spacefactor3000\relax}%
\providecommand \BibitemShut  [1]{\csname bibitem#1\endcsname}%
\let\auto@bib@innerbib\@empty
\bibitem [{\citenamefont {Brabec}\ and\ \citenamefont {Krausz}(2000)}]{Brabec:RMP00}%
  \BibitemOpen
  \bibfield  {author} {\bibinfo {author} {\bibfnamefont {T.}~\bibnamefont {Brabec}}\ and\ \bibinfo {author} {\bibfnamefont {F.}~\bibnamefont {Krausz}},\ }\href {\doibase 10.1103/RevModPhys.72.545} {\bibfield  {journal} {\bibinfo  {journal} {Rev. Mod. Phys.}\ }\textbf {\bibinfo {volume} {72}},\ \bibinfo {pages} {545} (\bibinfo {year} {2000})}\BibitemShut {NoStop}%
\bibitem [{\citenamefont {Corkum}\ and\ \citenamefont {Krausz}(2007)}]{Corkum:NatPhys17}%
  \BibitemOpen
  \bibfield  {author} {\bibinfo {author} {\bibfnamefont {P.~B.}\ \bibnamefont {Corkum}}\ and\ \bibinfo {author} {\bibfnamefont {F.}~\bibnamefont {Krausz}},\ }\href@noop {} {\bibfield  {journal} {\bibinfo  {journal} {Nat. Phys.}\ }\textbf {\bibinfo {volume} {3}},\ \bibinfo {pages} {381} (\bibinfo {year} {2007})}\BibitemShut {NoStop}%
\bibitem [{\citenamefont {Ramasesha}\ \emph {et~al.}(2016)\citenamefont {Ramasesha}, \citenamefont {Leone},\ and\ \citenamefont {Neumark}}]{Ramasesha:ARPC16}%
  \BibitemOpen
  \bibfield  {author} {\bibinfo {author} {\bibfnamefont {K.}~\bibnamefont {Ramasesha}}, \bibinfo {author} {\bibfnamefont {S.~R.}\ \bibnamefont {Leone}}, \ and\ \bibinfo {author} {\bibfnamefont {D.~M.}\ \bibnamefont {Neumark}},\ }\href {\doibase https://doi.org/10.1146/annurev-physchem-040215-112025} {\bibfield  {journal} {\bibinfo  {journal} {Annu. Rev. Phys. Chem.}\ }\textbf {\bibinfo {volume} {67}},\ \bibinfo {pages} {41} (\bibinfo {year} {2016})}\BibitemShut {NoStop}%
\bibitem [{\citenamefont {Nisoli}\ \emph {et~al.}(2017)\citenamefont {Nisoli}, \citenamefont {Decleva}, \citenamefont {Calegari}, \citenamefont {Palacios},\ and\ \citenamefont {Mart{\'\i}n}}]{Nisoli:ChemRev17}%
  \BibitemOpen
  \bibfield  {author} {\bibinfo {author} {\bibfnamefont {M.}~\bibnamefont {Nisoli}}, \bibinfo {author} {\bibfnamefont {P.}~\bibnamefont {Decleva}}, \bibinfo {author} {\bibfnamefont {F.}~\bibnamefont {Calegari}}, \bibinfo {author} {\bibfnamefont {A.}~\bibnamefont {Palacios}}, \ and\ \bibinfo {author} {\bibfnamefont {F.}~\bibnamefont {Mart{\'\i}n}},\ }\href@noop {} {\bibfield  {journal} {\bibinfo  {journal} {Chem. Rev.}\ }\textbf {\bibinfo {volume} {117}},\ \bibinfo {pages} {10760} (\bibinfo {year} {2017})}\BibitemShut {NoStop}%
\bibitem [{\citenamefont {Li}\ \emph {et~al.}(2020{\natexlab{a}})\citenamefont {Li}, \citenamefont {Lu}, \citenamefont {Chew}, \citenamefont {Han}, \citenamefont {Li}, \citenamefont {Wu}, \citenamefont {Wang}, \citenamefont {Ghimire},\ and\ \citenamefont {Chang}}]{Li:NatCom20}%
  \BibitemOpen
  \bibfield  {author} {\bibinfo {author} {\bibfnamefont {J.}~\bibnamefont {Li}}, \bibinfo {author} {\bibfnamefont {J.}~\bibnamefont {Lu}}, \bibinfo {author} {\bibfnamefont {A.}~\bibnamefont {Chew}}, \bibinfo {author} {\bibfnamefont {S.}~\bibnamefont {Han}}, \bibinfo {author} {\bibfnamefont {J.}~\bibnamefont {Li}}, \bibinfo {author} {\bibfnamefont {Y.}~\bibnamefont {Wu}}, \bibinfo {author} {\bibfnamefont {H.}~\bibnamefont {Wang}}, \bibinfo {author} {\bibfnamefont {S.}~\bibnamefont {Ghimire}}, \ and\ \bibinfo {author} {\bibfnamefont {Z.}~\bibnamefont {Chang}},\ }\href@noop {} {\bibfield  {journal} {\bibinfo  {journal} {Nat. Commun.}\ }\textbf {\bibinfo {volume} {11}},\ \bibinfo {pages} {2748} (\bibinfo {year} {2020}{\natexlab{a}})}\BibitemShut {NoStop}%
\bibitem [{\citenamefont {Itatani}\ \emph {et~al.}(2004)\citenamefont {Itatani}, \citenamefont {Levesque}, \citenamefont {Zeidler}, \citenamefont {Niikura}, \citenamefont {P{\'e}pin}, \citenamefont {Kieffer}, \citenamefont {Corkum},\ and\ \citenamefont {Villeneuve}}]{Itatani:Nat04}%
  \BibitemOpen
  \bibfield  {author} {\bibinfo {author} {\bibfnamefont {J.}~\bibnamefont {Itatani}}, \bibinfo {author} {\bibfnamefont {J.}~\bibnamefont {Levesque}}, \bibinfo {author} {\bibfnamefont {D.}~\bibnamefont {Zeidler}}, \bibinfo {author} {\bibfnamefont {H.}~\bibnamefont {Niikura}}, \bibinfo {author} {\bibfnamefont {H.}~\bibnamefont {P{\'e}pin}}, \bibinfo {author} {\bibfnamefont {J.-C.}\ \bibnamefont {Kieffer}}, \bibinfo {author} {\bibfnamefont {P.~B.}\ \bibnamefont {Corkum}}, \ and\ \bibinfo {author} {\bibfnamefont {D.~M.}\ \bibnamefont {Villeneuve}},\ }\href@noop {} {\bibfield  {journal} {\bibinfo  {journal} {Nature}\ }\textbf {\bibinfo {volume} {432}},\ \bibinfo {pages} {867} (\bibinfo {year} {2004})}\BibitemShut {NoStop}%
\bibitem [{\citenamefont {Haessler}\ \emph {et~al.}(2010)\citenamefont {Haessler}, \citenamefont {Caillat}, \citenamefont {Boutu}, \citenamefont {Giovanetti-Teixeira}, \citenamefont {Ruchon}, \citenamefont {Auguste}, \citenamefont {Diveki}, \citenamefont {Breger}, \citenamefont {Maquet}, \citenamefont {Carr{\'e}}, \citenamefont {Ta{\"i}eb},\ and\ \citenamefont {Salières}}]{Haessler:NatPhys10}%
  \BibitemOpen
  \bibfield  {author} {\bibinfo {author} {\bibfnamefont {S.}~\bibnamefont {Haessler}}, \bibinfo {author} {\bibfnamefont {J.}~\bibnamefont {Caillat}}, \bibinfo {author} {\bibfnamefont {W.}~\bibnamefont {Boutu}}, \bibinfo {author} {\bibfnamefont {C.}~\bibnamefont {Giovanetti-Teixeira}}, \bibinfo {author} {\bibfnamefont {T.}~\bibnamefont {Ruchon}}, \bibinfo {author} {\bibfnamefont {T.}~\bibnamefont {Auguste}}, \bibinfo {author} {\bibfnamefont {Z.}~\bibnamefont {Diveki}}, \bibinfo {author} {\bibfnamefont {P.}~\bibnamefont {Breger}}, \bibinfo {author} {\bibfnamefont {A.}~\bibnamefont {Maquet}}, \bibinfo {author} {\bibfnamefont {B.}~\bibnamefont {Carr{\'e}}}, \bibinfo {author} {\bibfnamefont {R.}~\bibnamefont {Ta{\"i}eb}}, \ and\ \bibinfo {author} {\bibfnamefont {P.}~\bibnamefont {Salières}},\ }\href@noop {} {\bibfield  {journal} {\bibinfo  {journal} {Nat. Phys.}\ }\textbf {\bibinfo {volume} {6}},\ \bibinfo {pages} {200} (\bibinfo {year} {2010})}\BibitemShut {NoStop}%
\bibitem [{\citenamefont {Lan}\ \emph {et~al.}(2017)\citenamefont {Lan}, \citenamefont {Ruhmann}, \citenamefont {He}, \citenamefont {Zhai}, \citenamefont {Wang}, \citenamefont {Zhu}, \citenamefont {Zhang}, \citenamefont {Zhou}, \citenamefont {Li}, \citenamefont {Lein},\ and\ \citenamefont {Lu}}]{Lan:prl17}%
  \BibitemOpen
  \bibfield  {author} {\bibinfo {author} {\bibfnamefont {P.}~\bibnamefont {Lan}}, \bibinfo {author} {\bibfnamefont {M.}~\bibnamefont {Ruhmann}}, \bibinfo {author} {\bibfnamefont {L.}~\bibnamefont {He}}, \bibinfo {author} {\bibfnamefont {C.}~\bibnamefont {Zhai}}, \bibinfo {author} {\bibfnamefont {F.}~\bibnamefont {Wang}}, \bibinfo {author} {\bibfnamefont {X.}~\bibnamefont {Zhu}}, \bibinfo {author} {\bibfnamefont {Q.}~\bibnamefont {Zhang}}, \bibinfo {author} {\bibfnamefont {Y.}~\bibnamefont {Zhou}}, \bibinfo {author} {\bibfnamefont {M.}~\bibnamefont {Li}}, \bibinfo {author} {\bibfnamefont {M.}~\bibnamefont {Lein}}, \ and\ \bibinfo {author} {\bibfnamefont {P.}~\bibnamefont {Lu}},\ }\href@noop {} {\bibfield  {journal} {\bibinfo  {journal} {Phys. Rev. Lett.}\ }\textbf {\bibinfo {volume} {119}},\ \bibinfo {pages} {033201} (\bibinfo {year} {2017})}\BibitemShut {NoStop}%
\bibitem [{\citenamefont {Kim}\ \emph {et~al.}(2014)\citenamefont {Kim}, \citenamefont {Villeneuve},\ and\ \citenamefont {Corkum}}]{Kim:NatPho14}%
  \BibitemOpen
  \bibfield  {author} {\bibinfo {author} {\bibfnamefont {K.~T.}\ \bibnamefont {Kim}}, \bibinfo {author} {\bibfnamefont {D.}~\bibnamefont {Villeneuve}}, \ and\ \bibinfo {author} {\bibfnamefont {P.}~\bibnamefont {Corkum}},\ }\href@noop {} {\bibfield  {journal} {\bibinfo  {journal} {Nat. Photonics.}\ }\textbf {\bibinfo {volume} {8}},\ \bibinfo {pages} {187} (\bibinfo {year} {2014})}\BibitemShut {NoStop}%
\bibitem [{\citenamefont {Peng}\ \emph {et~al.}(2019)\citenamefont {Peng}, \citenamefont {Marceau},\ and\ \citenamefont {Villeneuve}}]{Peng:NatRevPhys:19}%
  \BibitemOpen
  \bibfield  {author} {\bibinfo {author} {\bibfnamefont {P.}~\bibnamefont {Peng}}, \bibinfo {author} {\bibfnamefont {C.}~\bibnamefont {Marceau}}, \ and\ \bibinfo {author} {\bibfnamefont {D.~M.}\ \bibnamefont {Villeneuve}},\ }\href@noop {} {\bibfield  {journal} {\bibinfo  {journal} {Nat. Rev. Phys.}\ }\textbf {\bibinfo {volume} {1}},\ \bibinfo {pages} {144} (\bibinfo {year} {2019})}\BibitemShut {NoStop}%
\bibitem [{\citenamefont {Corkum}(1993)}]{Corkum:prl93}%
  \BibitemOpen
  \bibfield  {author} {\bibinfo {author} {\bibfnamefont {P.~B.}\ \bibnamefont {Corkum}},\ }\href {\doibase 10.1103/PhysRevLett.71.1994} {\bibfield  {journal} {\bibinfo  {journal} {Phys. Rev. Lett.}\ }\textbf {\bibinfo {volume} {71}},\ \bibinfo {pages} {1994} (\bibinfo {year} {1993})}\BibitemShut {NoStop}%
\bibitem [{\citenamefont {Lewenstein}\ \emph {et~al.}(1994)\citenamefont {Lewenstein}, \citenamefont {Balcou}, \citenamefont {Ivanov}, \citenamefont {L'Huillier},\ and\ \citenamefont {Corkum}}]{Lewenstein:pra94}%
  \BibitemOpen
  \bibfield  {author} {\bibinfo {author} {\bibfnamefont {M.}~\bibnamefont {Lewenstein}}, \bibinfo {author} {\bibfnamefont {P.}~\bibnamefont {Balcou}}, \bibinfo {author} {\bibfnamefont {M.~Y.}\ \bibnamefont {Ivanov}}, \bibinfo {author} {\bibfnamefont {A.}~\bibnamefont {L'Huillier}}, \ and\ \bibinfo {author} {\bibfnamefont {P.~B.}\ \bibnamefont {Corkum}},\ }\href {\doibase 10.1103/PhysRevA.49.2117} {\bibfield  {journal} {\bibinfo  {journal} {Phys. Rev. A}\ }\textbf {\bibinfo {volume} {49}},\ \bibinfo {pages} {2117} (\bibinfo {year} {1994})}\BibitemShut {NoStop}%
\bibitem [{\citenamefont {Mairesse}\ \emph {et~al.}(2003)\citenamefont {Mairesse}, \citenamefont {de~Bohan}, \citenamefont {Frasinski}, \citenamefont {Merdji}, \citenamefont {Dinu}, \citenamefont {Monchicourt}, \citenamefont {Breger}, \citenamefont {Kovačev}, \citenamefont {Taïeb}, \citenamefont {Carré}, \citenamefont {Muller}, \citenamefont {Agostini},\ and\ \citenamefont {Salières}}]{Mairesse:Sc03}%
  \BibitemOpen
  \bibfield  {author} {\bibinfo {author} {\bibfnamefont {Y.}~\bibnamefont {Mairesse}}, \bibinfo {author} {\bibfnamefont {A.}~\bibnamefont {de~Bohan}}, \bibinfo {author} {\bibfnamefont {L.~J.}\ \bibnamefont {Frasinski}}, \bibinfo {author} {\bibfnamefont {H.}~\bibnamefont {Merdji}}, \bibinfo {author} {\bibfnamefont {L.~C.}\ \bibnamefont {Dinu}}, \bibinfo {author} {\bibfnamefont {P.}~\bibnamefont {Monchicourt}}, \bibinfo {author} {\bibfnamefont {P.}~\bibnamefont {Breger}}, \bibinfo {author} {\bibfnamefont {M.}~\bibnamefont {Kovačev}}, \bibinfo {author} {\bibfnamefont {R.}~\bibnamefont {Taïeb}}, \bibinfo {author} {\bibfnamefont {B.}~\bibnamefont {Carré}}, \bibinfo {author} {\bibfnamefont {H.~G.}\ \bibnamefont {Muller}}, \bibinfo {author} {\bibfnamefont {P.}~\bibnamefont {Agostini}}, \ and\ \bibinfo {author} {\bibfnamefont {P.}~\bibnamefont {Salières}},\ }\href {\doibase 10.1126/science.1090277} {\bibfield  {journal} {\bibinfo  {journal} {Science}\ }\textbf {\bibinfo {volume} {302}},\ \bibinfo {pages} {1540}
  (\bibinfo {year} {2003})}\BibitemShut {NoStop}%
\bibitem [{\citenamefont {Niikura}\ \emph {et~al.}(2005)\citenamefont {Niikura}, \citenamefont {Villeneuve},\ and\ \citenamefont {Corkum}}]{Niikura:prl03}%
  \BibitemOpen
  \bibfield  {author} {\bibinfo {author} {\bibfnamefont {H.}~\bibnamefont {Niikura}}, \bibinfo {author} {\bibfnamefont {D.~M.}\ \bibnamefont {Villeneuve}}, \ and\ \bibinfo {author} {\bibfnamefont {P.~B.}\ \bibnamefont {Corkum}},\ }\href {\doibase 10.1103/PhysRevLett.94.083003} {\bibfield  {journal} {\bibinfo  {journal} {Phys. Rev. Lett.}\ }\textbf {\bibinfo {volume} {94}},\ \bibinfo {pages} {083003} (\bibinfo {year} {2005})}\BibitemShut {NoStop}%
\bibitem [{\citenamefont {Ben-Tal}\ \emph {et~al.}(1993)\citenamefont {Ben-Tal}, \citenamefont {Moiseyev},\ and\ \citenamefont {Beswick}}]{Ben-Tal_1993}%
  \BibitemOpen
  \bibfield  {author} {\bibinfo {author} {\bibfnamefont {N.}~\bibnamefont {Ben-Tal}}, \bibinfo {author} {\bibfnamefont {N.}~\bibnamefont {Moiseyev}}, \ and\ \bibinfo {author} {\bibfnamefont {A.}~\bibnamefont {Beswick}},\ }\href {\doibase 10.1088/0953-4075/26/18/012} {\bibfield  {journal} {\bibinfo  {journal} {J. Phys. B: At. Mol. Opt. Phys.}\ }\textbf {\bibinfo {volume} {26}},\ \bibinfo {pages} {3017} (\bibinfo {year} {1993})}\BibitemShut {NoStop}%
\bibitem [{\citenamefont {Neufeld}\ \emph {et~al.}(2019)\citenamefont {Neufeld}, \citenamefont {Podolsky},\ and\ \citenamefont {Cohen}}]{Neufeld:NatCom19}%
  \BibitemOpen
  \bibfield  {author} {\bibinfo {author} {\bibfnamefont {O.}~\bibnamefont {Neufeld}}, \bibinfo {author} {\bibfnamefont {D.}~\bibnamefont {Podolsky}}, \ and\ \bibinfo {author} {\bibfnamefont {O.}~\bibnamefont {Cohen}},\ }\href@noop {} {\bibfield  {journal} {\bibinfo  {journal} {Nat. Commun.}\ }\textbf {\bibinfo {volume} {10}},\ \bibinfo {pages} {405} (\bibinfo {year} {2019})}\BibitemShut {NoStop}%
\bibitem [{\citenamefont {Frumker}\ \emph {et~al.}(2012)\citenamefont {Frumker}, \citenamefont {Kajumba}, \citenamefont {Bertrand}, \citenamefont {W{\"o}rner}, \citenamefont {Hebeisen}, \citenamefont {Hockett}, \citenamefont {Spanner}, \citenamefont {Patchkovskii}, \citenamefont {Paulus}, \citenamefont {Villeneuve}, \citenamefont {Naumov},\ and\ \citenamefont {Corkum}}]{Frumker:PRL12}%
  \BibitemOpen
  \bibfield  {author} {\bibinfo {author} {\bibfnamefont {E.}~\bibnamefont {Frumker}}, \bibinfo {author} {\bibfnamefont {N.}~\bibnamefont {Kajumba}}, \bibinfo {author} {\bibfnamefont {J.~B.}\ \bibnamefont {Bertrand}}, \bibinfo {author} {\bibfnamefont {H.~J.}\ \bibnamefont {W{\"o}rner}}, \bibinfo {author} {\bibfnamefont {C.~T.}\ \bibnamefont {Hebeisen}}, \bibinfo {author} {\bibfnamefont {P.}~\bibnamefont {Hockett}}, \bibinfo {author} {\bibfnamefont {M.}~\bibnamefont {Spanner}}, \bibinfo {author} {\bibfnamefont {S.}~\bibnamefont {Patchkovskii}}, \bibinfo {author} {\bibfnamefont {G.~G.}\ \bibnamefont {Paulus}}, \bibinfo {author} {\bibfnamefont {D.~M.}\ \bibnamefont {Villeneuve}}, \bibinfo {author} {\bibfnamefont {A.}~\bibnamefont {Naumov}}, \ and\ \bibinfo {author} {\bibfnamefont {P.~B.}\ \bibnamefont {Corkum}},\ }\href@noop {} {\bibfield  {journal} {\bibinfo  {journal} {Phys. Rev. Lett.}\ }\textbf {\bibinfo {volume} {109}},\ \bibinfo {pages} {233904} (\bibinfo {year} {2012})}\BibitemShut {NoStop}%
\bibitem [{\citenamefont {Kraus}\ \emph {et~al.}(2012)\citenamefont {Kraus}, \citenamefont {Rupenyan},\ and\ \citenamefont {W\"orner}}]{Kraus:prl12}%
  \BibitemOpen
  \bibfield  {author} {\bibinfo {author} {\bibfnamefont {P.~M.}\ \bibnamefont {Kraus}}, \bibinfo {author} {\bibfnamefont {A.}~\bibnamefont {Rupenyan}}, \ and\ \bibinfo {author} {\bibfnamefont {H.~J.}\ \bibnamefont {W\"orner}},\ }\href {\doibase 10.1103/PhysRevLett.109.233903} {\bibfield  {journal} {\bibinfo  {journal} {Phys. Rev. Lett.}\ }\textbf {\bibinfo {volume} {109}},\ \bibinfo {pages} {233903} (\bibinfo {year} {2012})}\BibitemShut {NoStop}%
\bibitem [{\citenamefont {Chen}\ \emph {et~al.}(2013)\citenamefont {Chen}, \citenamefont {Fu},\ and\ \citenamefont {Liu}}]{Chen:prl13}%
  \BibitemOpen
  \bibfield  {author} {\bibinfo {author} {\bibfnamefont {Y.~J.}\ \bibnamefont {Chen}}, \bibinfo {author} {\bibfnamefont {L.~B.}\ \bibnamefont {Fu}}, \ and\ \bibinfo {author} {\bibfnamefont {J.}~\bibnamefont {Liu}},\ }\href {\doibase 10.1103/PhysRevLett.111.073902} {\bibfield  {journal} {\bibinfo  {journal} {Phys. Rev. Lett.}\ }\textbf {\bibinfo {volume} {111}},\ \bibinfo {pages} {073902} (\bibinfo {year} {2013})}\BibitemShut {NoStop}%
\bibitem [{\citenamefont {Nguyen-Huynh}\ \emph {et~al.}(2020)\citenamefont {Nguyen-Huynh}, \citenamefont {Le}, \citenamefont {Phan}, \citenamefont {Nguyen},\ and\ \citenamefont {Tran}}]{Nguyen:ComPhy20}%
  \BibitemOpen
  \bibfield  {author} {\bibinfo {author} {\bibfnamefont {K.-N.}\ \bibnamefont {Nguyen-Huynh}}, \bibinfo {author} {\bibfnamefont {C.-T.}\ \bibnamefont {Le}}, \bibinfo {author} {\bibfnamefont {N.-L.~T.}\ \bibnamefont {Phan}}, \bibinfo {author} {\bibfnamefont {H.~T.}\ \bibnamefont {Nguyen}}, \ and\ \bibinfo {author} {\bibfnamefont {L.-P.}\ \bibnamefont {Tran}},\ }\href {\doibase 10.15625/0868-3166/30/3/14865} {\bibfield  {journal} {\bibinfo  {journal} {Comm. Phys.}\ }\textbf {\bibinfo {volume} {30}},\ \bibinfo {pages} {197} (\bibinfo {year} {2020})}\BibitemShut {NoStop}%
\bibitem [{\citenamefont {Phan}\ \emph {et~al.}(2019)\citenamefont {Phan}, \citenamefont {Le}, \citenamefont {Hoang},\ and\ \citenamefont {Le}}]{Phan:PCCP19}%
  \BibitemOpen
  \bibfield  {author} {\bibinfo {author} {\bibfnamefont {N.-L.}\ \bibnamefont {Phan}}, \bibinfo {author} {\bibfnamefont {C.-T.}\ \bibnamefont {Le}}, \bibinfo {author} {\bibfnamefont {V.-H.}\ \bibnamefont {Hoang}}, \ and\ \bibinfo {author} {\bibfnamefont {V.-H.}\ \bibnamefont {Le}},\ }\href {\doibase 10.1039/C9CP04064A} {\bibfield  {journal} {\bibinfo  {journal} {Phys. Chem. Chem. Phys.}\ }\textbf {\bibinfo {volume} {21}},\ \bibinfo {pages} {24177} (\bibinfo {year} {2019})}\BibitemShut {NoStop}%
\bibitem [{\citenamefont {Nguyen}\ \emph {et~al.}(2022{\natexlab{a}})\citenamefont {Nguyen}, \citenamefont {Nguyen}, \citenamefont {Phan}, \citenamefont {Le}, \citenamefont {Vu}, \citenamefont {Tran},\ and\ \citenamefont {Le}}]{Nguyen:pra22}%
  \BibitemOpen
  \bibfield  {author} {\bibinfo {author} {\bibfnamefont {H.~T.}\ \bibnamefont {Nguyen}}, \bibinfo {author} {\bibfnamefont {K.-N.~H.}\ \bibnamefont {Nguyen}}, \bibinfo {author} {\bibfnamefont {N.-L.}\ \bibnamefont {Phan}}, \bibinfo {author} {\bibfnamefont {C.-T.}\ \bibnamefont {Le}}, \bibinfo {author} {\bibfnamefont {D.}~\bibnamefont {Vu}}, \bibinfo {author} {\bibfnamefont {L.-P.}\ \bibnamefont {Tran}}, \ and\ \bibinfo {author} {\bibfnamefont {V.-H.}\ \bibnamefont {Le}},\ }\href {\doibase 10.1103/PhysRevA.105.023106} {\bibfield  {journal} {\bibinfo  {journal} {Phys. Rev. A}\ }\textbf {\bibinfo {volume} {105}},\ \bibinfo {pages} {023106} (\bibinfo {year} {2022}{\natexlab{a}})}\BibitemShut {NoStop}%
\bibitem [{\citenamefont {Nguyen}\ \emph {et~al.}(2022{\natexlab{b}})\citenamefont {Nguyen}, \citenamefont {Phan}, \citenamefont {Le}, \citenamefont {Vu},\ and\ \citenamefont {Le}}]{Ngan:pra20}%
  \BibitemOpen
  \bibfield  {author} {\bibinfo {author} {\bibfnamefont {K.-N.~H.}\ \bibnamefont {Nguyen}}, \bibinfo {author} {\bibfnamefont {N.-L.}\ \bibnamefont {Phan}}, \bibinfo {author} {\bibfnamefont {C.-T.}\ \bibnamefont {Le}}, \bibinfo {author} {\bibfnamefont {D.}~\bibnamefont {Vu}}, \ and\ \bibinfo {author} {\bibfnamefont {V.-H.}\ \bibnamefont {Le}},\ }\href {\doibase 10.1103/PhysRevA.106.063108} {\bibfield  {journal} {\bibinfo  {journal} {Phys. Rev. A}\ }\textbf {\bibinfo {volume} {106}},\ \bibinfo {pages} {063108} (\bibinfo {year} {2022}{\natexlab{b}})}\BibitemShut {NoStop}%
\bibitem [{\citenamefont {Luu}\ and\ \citenamefont {W{\"o}rner}(2018)}]{Luu:NatCom18}%
  \BibitemOpen
  \bibfield  {author} {\bibinfo {author} {\bibfnamefont {T.~T.}\ \bibnamefont {Luu}}\ and\ \bibinfo {author} {\bibfnamefont {H.~J.}\ \bibnamefont {W{\"o}rner}},\ }\href@noop {} {\bibfield  {journal} {\bibinfo  {journal} {Nat. Commun.}\ }\textbf {\bibinfo {volume} {9}},\ \bibinfo {pages} {916} (\bibinfo {year} {2018})}\BibitemShut {NoStop}%
\bibitem [{\citenamefont {Uzan-Narovlansky}\ \emph {et~al.}(2023)\citenamefont {Uzan-Narovlansky}, \citenamefont {Orenstein}, \citenamefont {Shames}, \citenamefont {Even~Tzur}, \citenamefont {Kneller}, \citenamefont {Bruner}, \citenamefont {Arusi-Parpar}, \citenamefont {Cohen},\ and\ \citenamefont {Dudovich}}]{Uzan:prl23}%
  \BibitemOpen
  \bibfield  {author} {\bibinfo {author} {\bibfnamefont {A.~J.}\ \bibnamefont {Uzan-Narovlansky}}, \bibinfo {author} {\bibfnamefont {G.}~\bibnamefont {Orenstein}}, \bibinfo {author} {\bibfnamefont {S.}~\bibnamefont {Shames}}, \bibinfo {author} {\bibfnamefont {M.}~\bibnamefont {Even~Tzur}}, \bibinfo {author} {\bibfnamefont {O.}~\bibnamefont {Kneller}}, \bibinfo {author} {\bibfnamefont {B.~D.}\ \bibnamefont {Bruner}}, \bibinfo {author} {\bibfnamefont {T.}~\bibnamefont {Arusi-Parpar}}, \bibinfo {author} {\bibfnamefont {O.}~\bibnamefont {Cohen}}, \ and\ \bibinfo {author} {\bibfnamefont {N.}~\bibnamefont {Dudovich}},\ }\href {\doibase 10.1103/PhysRevLett.131.223802} {\bibfield  {journal} {\bibinfo  {journal} {Phys. Rev. Lett.}\ }\textbf {\bibinfo {volume} {131}},\ \bibinfo {pages} {223802} (\bibinfo {year} {2023})}\BibitemShut {NoStop}%
\bibitem [{\citenamefont {Li}\ \emph {et~al.}(2023)\citenamefont {Li}, \citenamefont {Tang}, \citenamefont {Ortmann}, \citenamefont {Talbert}, \citenamefont {Blaga}, \citenamefont {Lai}, \citenamefont {Wang}, \citenamefont {Cheng}, \citenamefont {Yang}, \citenamefont {Landsman}, \citenamefont {Agostini},\ and\ \citenamefont {DiMauro}}]{Li:NC23}%
  \BibitemOpen
  \bibfield  {author} {\bibinfo {author} {\bibfnamefont {S.}~\bibnamefont {Li}}, \bibinfo {author} {\bibfnamefont {Y.}~\bibnamefont {Tang}}, \bibinfo {author} {\bibfnamefont {L.}~\bibnamefont {Ortmann}}, \bibinfo {author} {\bibfnamefont {B.~K.}\ \bibnamefont {Talbert}}, \bibinfo {author} {\bibfnamefont {C.~I.}\ \bibnamefont {Blaga}}, \bibinfo {author} {\bibfnamefont {Y.~H.}\ \bibnamefont {Lai}}, \bibinfo {author} {\bibfnamefont {Z.}~\bibnamefont {Wang}}, \bibinfo {author} {\bibfnamefont {Y.}~\bibnamefont {Cheng}}, \bibinfo {author} {\bibfnamefont {F.}~\bibnamefont {Yang}}, \bibinfo {author} {\bibfnamefont {A.~S.}\ \bibnamefont {Landsman}}, \bibinfo {author} {\bibfnamefont {P.}~\bibnamefont {Agostini}}, \ and\ \bibinfo {author} {\bibfnamefont {L.~F.}\ \bibnamefont {DiMauro}},\ }\href@noop {} {\bibfield  {journal} {\bibinfo  {journal} {Nat. Commun.}\ }\textbf {\bibinfo {volume} {14}},\ \bibinfo {pages} {2603} (\bibinfo {year} {2023})}\BibitemShut {NoStop}%
\bibitem [{\citenamefont {Lambert}\ \emph {et~al.}(2015)\citenamefont {Lambert}, \citenamefont {Vodungbo}, \citenamefont {Gautier}, \citenamefont {Mahieu}, \citenamefont {Malka}, \citenamefont {Sebban}, \citenamefont {Zeitoun}, \citenamefont {Luning}, \citenamefont {Perron}, \citenamefont {Andreev}, \citenamefont {Stremoukhov}, \citenamefont {Ardana-Lamas}, \citenamefont {Dax}, \citenamefont {Hauri}, \citenamefont {Sardinha},\ and\ \citenamefont {Fajardo}}]{Lambert:NatCom15}%
  \BibitemOpen
  \bibfield  {author} {\bibinfo {author} {\bibfnamefont {G.}~\bibnamefont {Lambert}}, \bibinfo {author} {\bibfnamefont {B.}~\bibnamefont {Vodungbo}}, \bibinfo {author} {\bibfnamefont {J.}~\bibnamefont {Gautier}}, \bibinfo {author} {\bibfnamefont {B.}~\bibnamefont {Mahieu}}, \bibinfo {author} {\bibfnamefont {V.}~\bibnamefont {Malka}}, \bibinfo {author} {\bibfnamefont {S.}~\bibnamefont {Sebban}}, \bibinfo {author} {\bibfnamefont {P.}~\bibnamefont {Zeitoun}}, \bibinfo {author} {\bibfnamefont {J.}~\bibnamefont {Luning}}, \bibinfo {author} {\bibfnamefont {J.}~\bibnamefont {Perron}}, \bibinfo {author} {\bibfnamefont {A.}~\bibnamefont {Andreev}}, \bibinfo {author} {\bibfnamefont {S.}~\bibnamefont {Stremoukhov}}, \bibinfo {author} {\bibfnamefont {F.}~\bibnamefont {Ardana-Lamas}}, \bibinfo {author} {\bibfnamefont {A.}~\bibnamefont {Dax}}, \bibinfo {author} {\bibfnamefont {C.~P.}\ \bibnamefont {Hauri}}, \bibinfo {author} {\bibfnamefont {A.}~\bibnamefont {Sardinha}}, \ and\ \bibinfo {author} {\bibfnamefont
  {M.}~\bibnamefont {Fajardo}},\ }\href@noop {} {\bibfield  {journal} {\bibinfo  {journal} {Nat. Commun.}\ }\textbf {\bibinfo {volume} {6}},\ \bibinfo {pages} {6167} (\bibinfo {year} {2015})}\BibitemShut {NoStop}%
\bibitem [{\citenamefont {Vampa}\ \emph {et~al.}(2018)\citenamefont {Vampa}, \citenamefont {Hammond}, \citenamefont {Taucer}, \citenamefont {Ding}, \citenamefont {Ropagnol}, \citenamefont {Ozaki}, \citenamefont {Delprat}, \citenamefont {Chaker}, \citenamefont {Thir{\'e}}, \citenamefont {Schmidt}, \citenamefont {L{\'e}gar{\'e}}, \citenamefont {Klug}, \citenamefont {Naumov}, \citenamefont {Villeneuve}, \citenamefont {Staudte},\ and\ \citenamefont {Corkum}}]{Vampa:NP18}%
  \BibitemOpen
  \bibfield  {author} {\bibinfo {author} {\bibfnamefont {G.}~\bibnamefont {Vampa}}, \bibinfo {author} {\bibfnamefont {T.}~\bibnamefont {Hammond}}, \bibinfo {author} {\bibfnamefont {M.}~\bibnamefont {Taucer}}, \bibinfo {author} {\bibfnamefont {X.}~\bibnamefont {Ding}}, \bibinfo {author} {\bibfnamefont {X.}~\bibnamefont {Ropagnol}}, \bibinfo {author} {\bibfnamefont {T.}~\bibnamefont {Ozaki}}, \bibinfo {author} {\bibfnamefont {S.}~\bibnamefont {Delprat}}, \bibinfo {author} {\bibfnamefont {M.}~\bibnamefont {Chaker}}, \bibinfo {author} {\bibfnamefont {N.}~\bibnamefont {Thir{\'e}}}, \bibinfo {author} {\bibfnamefont {B.}~\bibnamefont {Schmidt}}, \bibinfo {author} {\bibfnamefont {F.}~\bibnamefont {L{\'e}gar{\'e}}}, \bibinfo {author} {\bibfnamefont {D.~D.}\ \bibnamefont {Klug}}, \bibinfo {author} {\bibfnamefont {A.~Y.}\ \bibnamefont {Naumov}}, \bibinfo {author} {\bibfnamefont {D.~M.}\ \bibnamefont {Villeneuve}}, \bibinfo {author} {\bibfnamefont {A.}~\bibnamefont {Staudte}}, \ and\ \bibinfo {author} {\bibfnamefont
  {P.~B.}\ \bibnamefont {Corkum}},\ }\href@noop {} {\bibfield  {journal} {\bibinfo  {journal} {Nat. Photonics}\ }\textbf {\bibinfo {volume} {12}},\ \bibinfo {pages} {465} (\bibinfo {year} {2018})}\BibitemShut {NoStop}%
\bibitem [{\citenamefont {Luu}\ and\ \citenamefont {W\"orner}(2018)}]{Luu:pra18}%
  \BibitemOpen
  \bibfield  {author} {\bibinfo {author} {\bibfnamefont {T.~T.}\ \bibnamefont {Luu}}\ and\ \bibinfo {author} {\bibfnamefont {H.~J.}\ \bibnamefont {W\"orner}},\ }\href {\doibase 10.1103/PhysRevA.98.041802} {\bibfield  {journal} {\bibinfo  {journal} {Phys. Rev. A}\ }\textbf {\bibinfo {volume} {98}},\ \bibinfo {pages} {041802} (\bibinfo {year} {2018})}\BibitemShut {NoStop}%
\bibitem [{\citenamefont {Shafir}\ \emph {et~al.}(2009)\citenamefont {Shafir}, \citenamefont {Mairesse}, \citenamefont {Villeneuve}, \citenamefont {Corkum},\ and\ \citenamefont {Dudovich}}]{Shafir:NatPhys09}%
  \BibitemOpen
  \bibfield  {author} {\bibinfo {author} {\bibfnamefont {D.}~\bibnamefont {Shafir}}, \bibinfo {author} {\bibfnamefont {Y.}~\bibnamefont {Mairesse}}, \bibinfo {author} {\bibfnamefont {D.}~\bibnamefont {Villeneuve}}, \bibinfo {author} {\bibfnamefont {P.}~\bibnamefont {Corkum}}, \ and\ \bibinfo {author} {\bibfnamefont {N.}~\bibnamefont {Dudovich}},\ }\href@noop {} {\bibfield  {journal} {\bibinfo  {journal} {Nat. Phys.}\ }\textbf {\bibinfo {volume} {5}},\ \bibinfo {pages} {412} (\bibinfo {year} {2009})}\BibitemShut {NoStop}%
\bibitem [{\citenamefont {Niikura}\ \emph {et~al.}(2010)\citenamefont {Niikura}, \citenamefont {Dudovich}, \citenamefont {Villeneuve},\ and\ \citenamefont {Corkum}}]{Niikura:prl10}%
  \BibitemOpen
  \bibfield  {author} {\bibinfo {author} {\bibfnamefont {H.}~\bibnamefont {Niikura}}, \bibinfo {author} {\bibfnamefont {N.}~\bibnamefont {Dudovich}}, \bibinfo {author} {\bibfnamefont {D.~M.}\ \bibnamefont {Villeneuve}}, \ and\ \bibinfo {author} {\bibfnamefont {P.~B.}\ \bibnamefont {Corkum}},\ }\href {\doibase 10.1103/PhysRevLett.105.053003} {\bibfield  {journal} {\bibinfo  {journal} {Phys. Rev. Lett.}\ }\textbf {\bibinfo {volume} {105}},\ \bibinfo {pages} {053003} (\bibinfo {year} {2010})}\BibitemShut {NoStop}%
\bibitem [{\citenamefont {Li}\ \emph {et~al.}(2020{\natexlab{b}})\citenamefont {Li}, \citenamefont {Zhang}, \citenamefont {Zhang}, \citenamefont {Yan},\ and\ \citenamefont {Jiang}}]{Li:pra20}%
  \BibitemOpen
  \bibfield  {author} {\bibinfo {author} {\bibfnamefont {B.-Y.}\ \bibnamefont {Li}}, \bibinfo {author} {\bibfnamefont {J.}~\bibnamefont {Zhang}}, \bibinfo {author} {\bibfnamefont {Y.}~\bibnamefont {Zhang}}, \bibinfo {author} {\bibfnamefont {T.-M.}\ \bibnamefont {Yan}}, \ and\ \bibinfo {author} {\bibfnamefont {Y.}~\bibnamefont {Jiang}},\ }\href@noop {} {\bibfield  {journal} {\bibinfo  {journal} {Phys. Rev. A}\ }\textbf {\bibinfo {volume} {102}},\ \bibinfo {pages} {063102} (\bibinfo {year} {2020}{\natexlab{b}})}\BibitemShut {NoStop}%
\bibitem [{\citenamefont {Trieu}\ \emph {et~al.}(2023)\citenamefont {Trieu}, \citenamefont {Phan}, \citenamefont {Truong}, \citenamefont {Nguyen}, \citenamefont {Le}, \citenamefont {Vu},\ and\ \citenamefont {Le}}]{Trieu:PRA23}%
  \BibitemOpen
  \bibfield  {author} {\bibinfo {author} {\bibfnamefont {D.-A.}\ \bibnamefont {Trieu}}, \bibinfo {author} {\bibfnamefont {N.-L.}\ \bibnamefont {Phan}}, \bibinfo {author} {\bibfnamefont {Q.-H.}\ \bibnamefont {Truong}}, \bibinfo {author} {\bibfnamefont {H.~T.}\ \bibnamefont {Nguyen}}, \bibinfo {author} {\bibfnamefont {C.-T.}\ \bibnamefont {Le}}, \bibinfo {author} {\bibfnamefont {D.}~\bibnamefont {Vu}}, \ and\ \bibinfo {author} {\bibfnamefont {V.-H.}\ \bibnamefont {Le}},\ }\href {\doibase 10.1103/PhysRevA.108.023109} {\bibfield  {journal} {\bibinfo  {journal} {Phys. Rev. A}\ }\textbf {\bibinfo {volume} {108}},\ \bibinfo {pages} {023109} (\bibinfo {year} {2023})}\BibitemShut {NoStop}%
\bibitem [{\citenamefont {Paulus}\ \emph {et~al.}(2003)\citenamefont {Paulus}, \citenamefont {Lindner}, \citenamefont {Walther}, \citenamefont {Baltu\ifmmode~\check{s}\else \v{s}\fi{}ka}, \citenamefont {Goulielmakis}, \citenamefont {Lezius},\ and\ \citenamefont {Krausz}}]{Paulus:prl03}%
  \BibitemOpen
  \bibfield  {author} {\bibinfo {author} {\bibfnamefont {G.~G.}\ \bibnamefont {Paulus}}, \bibinfo {author} {\bibfnamefont {F.}~\bibnamefont {Lindner}}, \bibinfo {author} {\bibfnamefont {H.}~\bibnamefont {Walther}}, \bibinfo {author} {\bibfnamefont {A.}~\bibnamefont {Baltu\ifmmode~\check{s}\else \v{s}\fi{}ka}}, \bibinfo {author} {\bibfnamefont {E.}~\bibnamefont {Goulielmakis}}, \bibinfo {author} {\bibfnamefont {M.}~\bibnamefont {Lezius}}, \ and\ \bibinfo {author} {\bibfnamefont {F.}~\bibnamefont {Krausz}},\ }\href {\doibase 10.1103/PhysRevLett.91.253004} {\bibfield  {journal} {\bibinfo  {journal} {Phys. Rev. Lett.}\ }\textbf {\bibinfo {volume} {91}},\ \bibinfo {pages} {253004} (\bibinfo {year} {2003})}\BibitemShut {NoStop}%
\bibitem [{\citenamefont {Wittmann}\ \emph {et~al.}(2009)\citenamefont {Wittmann}, \citenamefont {Horvath}, \citenamefont {Helml}, \citenamefont {Sch{\"a}tzel}, \citenamefont {Gu}, \citenamefont {Cavalieri}, \citenamefont {Paulus},\ and\ \citenamefont {Kienberger}}]{Wittmann:NatPhys09}%
  \BibitemOpen
  \bibfield  {author} {\bibinfo {author} {\bibfnamefont {T.}~\bibnamefont {Wittmann}}, \bibinfo {author} {\bibfnamefont {B.}~\bibnamefont {Horvath}}, \bibinfo {author} {\bibfnamefont {W.}~\bibnamefont {Helml}}, \bibinfo {author} {\bibfnamefont {M.~G.}\ \bibnamefont {Sch{\"a}tzel}}, \bibinfo {author} {\bibfnamefont {X.}~\bibnamefont {Gu}}, \bibinfo {author} {\bibfnamefont {A.~L.}\ \bibnamefont {Cavalieri}}, \bibinfo {author} {\bibfnamefont {G.}~\bibnamefont {Paulus}}, \ and\ \bibinfo {author} {\bibfnamefont {R.}~\bibnamefont {Kienberger}},\ }\href@noop {} {\bibfield  {journal} {\bibinfo  {journal} {Nat. Phys.}\ }\textbf {\bibinfo {volume} {5}},\ \bibinfo {pages} {357} (\bibinfo {year} {2009})}\BibitemShut {NoStop}%
\bibitem [{\citenamefont {Hauri}\ \emph {et~al.}(2004)\citenamefont {Hauri}, \citenamefont {Kornelis}, \citenamefont {Helbing}, \citenamefont {Heinrich}, \citenamefont {Couairon}, \citenamefont {Mysyrowicz}, \citenamefont {Biegert},\ and\ \citenamefont {Keller}}]{Hauri:aplB04}%
  \BibitemOpen
  \bibfield  {author} {\bibinfo {author} {\bibfnamefont {C.~P.}\ \bibnamefont {Hauri}}, \bibinfo {author} {\bibfnamefont {W.}~\bibnamefont {Kornelis}}, \bibinfo {author} {\bibfnamefont {F.}~\bibnamefont {Helbing}}, \bibinfo {author} {\bibfnamefont {A.}~\bibnamefont {Heinrich}}, \bibinfo {author} {\bibfnamefont {A.}~\bibnamefont {Couairon}}, \bibinfo {author} {\bibfnamefont {A.}~\bibnamefont {Mysyrowicz}}, \bibinfo {author} {\bibfnamefont {J.}~\bibnamefont {Biegert}}, \ and\ \bibinfo {author} {\bibfnamefont {U.}~\bibnamefont {Keller}},\ }\href@noop {} {\bibfield  {journal} {\bibinfo  {journal} {Appl. Phys. B}\ }\textbf {\bibinfo {volume} {79}},\ \bibinfo {pages} {673} (\bibinfo {year} {2004})}\BibitemShut {NoStop}%
\bibitem [{\citenamefont {Naumov}\ \emph {et~al.}(2015)\citenamefont {Naumov}, \citenamefont {Villeneuve},\ and\ \citenamefont {Niikura}}]{Naumov:pra15}%
  \BibitemOpen
  \bibfield  {author} {\bibinfo {author} {\bibfnamefont {A.~Y.}\ \bibnamefont {Naumov}}, \bibinfo {author} {\bibfnamefont {D.}~\bibnamefont {Villeneuve}}, \ and\ \bibinfo {author} {\bibfnamefont {H.}~\bibnamefont {Niikura}},\ }\href@noop {} {\bibfield  {journal} {\bibinfo  {journal} {Phys. Rev. A}\ }\textbf {\bibinfo {volume} {91}},\ \bibinfo {pages} {063421} (\bibinfo {year} {2015})}\BibitemShut {NoStop}%
\bibitem [{\citenamefont {Hollinger}\ \emph {et~al.}(2020)\citenamefont {Hollinger}, \citenamefont {Hoff}, \citenamefont {Wustelt}, \citenamefont {Skruszewicz}, \citenamefont {Zhang}, \citenamefont {Kang}, \citenamefont {W\"{u}rzler}, \citenamefont {Jungnickel}, \citenamefont {Dumergue}, \citenamefont {Nayak}, \citenamefont {Flender}, \citenamefont {Haizer}, \citenamefont {Kurucz}, \citenamefont {Kiss}, \citenamefont {K\"{u}hn}, \citenamefont {Cormier}, \citenamefont {Spielmann}, \citenamefont {Paulus}, \citenamefont {Tzallas},\ and\ \citenamefont {K\"{u}bel}}]{Hollinger:oe20}%
  \BibitemOpen
  \bibfield  {author} {\bibinfo {author} {\bibfnamefont {R.}~\bibnamefont {Hollinger}}, \bibinfo {author} {\bibfnamefont {D.}~\bibnamefont {Hoff}}, \bibinfo {author} {\bibfnamefont {P.}~\bibnamefont {Wustelt}}, \bibinfo {author} {\bibfnamefont {S.}~\bibnamefont {Skruszewicz}}, \bibinfo {author} {\bibfnamefont {Y.}~\bibnamefont {Zhang}}, \bibinfo {author} {\bibfnamefont {H.}~\bibnamefont {Kang}}, \bibinfo {author} {\bibfnamefont {D.}~\bibnamefont {W\"{u}rzler}}, \bibinfo {author} {\bibfnamefont {T.}~\bibnamefont {Jungnickel}}, \bibinfo {author} {\bibfnamefont {M.}~\bibnamefont {Dumergue}}, \bibinfo {author} {\bibfnamefont {A.}~\bibnamefont {Nayak}}, \bibinfo {author} {\bibfnamefont {R.}~\bibnamefont {Flender}}, \bibinfo {author} {\bibfnamefont {L.}~\bibnamefont {Haizer}}, \bibinfo {author} {\bibfnamefont {M.}~\bibnamefont {Kurucz}}, \bibinfo {author} {\bibfnamefont {B.}~\bibnamefont {Kiss}}, \bibinfo {author} {\bibfnamefont {S.}~\bibnamefont {K\"{u}hn}}, \bibinfo {author} {\bibfnamefont {E.}~\bibnamefont
  {Cormier}}, \bibinfo {author} {\bibfnamefont {C.}~\bibnamefont {Spielmann}}, \bibinfo {author} {\bibfnamefont {G.~G.}\ \bibnamefont {Paulus}}, \bibinfo {author} {\bibfnamefont {P.}~\bibnamefont {Tzallas}}, \ and\ \bibinfo {author} {\bibfnamefont {M.}~\bibnamefont {K\"{u}bel}},\ }\href {\doibase 10.1364/OE.383484} {\bibfield  {journal} {\bibinfo  {journal} {Opt. Express}\ }\textbf {\bibinfo {volume} {28}},\ \bibinfo {pages} {7314} (\bibinfo {year} {2020})}\BibitemShut {NoStop}%
\bibitem [{\citenamefont {You}\ \emph {et~al.}(2017)\citenamefont {You}, \citenamefont {Wu}, \citenamefont {Yin}, \citenamefont {Chew}, \citenamefont {Ren}, \citenamefont {Gholam-Mirzaei}, \citenamefont {Browne}, \citenamefont {Chini}, \citenamefont {Chang}, \citenamefont {Schafer}, \citenamefont {Gaarde},\ and\ \citenamefont {Ghimire}}]{You:OL17}%
  \BibitemOpen
  \bibfield  {author} {\bibinfo {author} {\bibfnamefont {Y.~S.}\ \bibnamefont {You}}, \bibinfo {author} {\bibfnamefont {M.}~\bibnamefont {Wu}}, \bibinfo {author} {\bibfnamefont {Y.}~\bibnamefont {Yin}}, \bibinfo {author} {\bibfnamefont {A.}~\bibnamefont {Chew}}, \bibinfo {author} {\bibfnamefont {X.}~\bibnamefont {Ren}}, \bibinfo {author} {\bibfnamefont {S.}~\bibnamefont {Gholam-Mirzaei}}, \bibinfo {author} {\bibfnamefont {D.~A.}\ \bibnamefont {Browne}}, \bibinfo {author} {\bibfnamefont {M.}~\bibnamefont {Chini}}, \bibinfo {author} {\bibfnamefont {Z.}~\bibnamefont {Chang}}, \bibinfo {author} {\bibfnamefont {K.~J.}\ \bibnamefont {Schafer}}, \bibinfo {author} {\bibfnamefont {M.~B.}\ \bibnamefont {Gaarde}}, \ and\ \bibinfo {author} {\bibfnamefont {S.}~\bibnamefont {Ghimire}},\ }\href@noop {} {\bibfield  {journal} {\bibinfo  {journal} {Opt. Lett.}\ }\textbf {\bibinfo {volume} {42}},\ \bibinfo {pages} {1816} (\bibinfo {year} {2017})}\BibitemShut {NoStop}%
\bibitem [{\citenamefont {Trieu}\ \emph {et~al.}(2024)\citenamefont {Trieu}, \citenamefont {Nguyen}, \citenamefont {Nguyen}, \citenamefont {Tran}, \citenamefont {Le},\ and\ \citenamefont {Phan}}]{Trieu:24}%
  \BibitemOpen
  \bibfield  {author} {\bibinfo {author} {\bibfnamefont {D.-A.}\ \bibnamefont {Trieu}}, \bibinfo {author} {\bibfnamefont {T.-T.~D.}\ \bibnamefont {Nguyen}}, \bibinfo {author} {\bibfnamefont {T.-D.~D.}\ \bibnamefont {Nguyen}}, \bibinfo {author} {\bibfnamefont {T.}~\bibnamefont {Tran}}, \bibinfo {author} {\bibfnamefont {V.-H.}\ \bibnamefont {Le}}, \ and\ \bibinfo {author} {\bibfnamefont {N.-L.}\ \bibnamefont {Phan}},\ }\href@noop {} {\enquote {\bibinfo {title} {Analytically controlling laser-induced electron phase in sub-cycle motion},}\ } (\bibinfo {year} {2024}),\ \Eprint {http://arxiv.org/abs/2405.11753} {arXiv:2405.11753 [physics.optics]} \BibitemShut {NoStop}%
\bibitem [{\citenamefont {Bian}\ and\ \citenamefont {Bandrauk}(2014)}]{Bian:prl14}%
  \BibitemOpen
  \bibfield  {author} {\bibinfo {author} {\bibfnamefont {X.-B.}\ \bibnamefont {Bian}}\ and\ \bibinfo {author} {\bibfnamefont {A.~D.}\ \bibnamefont {Bandrauk}},\ }\href {\doibase 10.1103/PhysRevLett.113.193901} {\bibfield  {journal} {\bibinfo  {journal} {Phys. Rev. Lett.}\ }\textbf {\bibinfo {volume} {113}},\ \bibinfo {pages} {193901} (\bibinfo {year} {2014})}\BibitemShut {NoStop}%
\bibitem [{\citenamefont {Zhou}\ \emph {et~al.}(1996)\citenamefont {Zhou}, \citenamefont {Peatross}, \citenamefont {Murnane}, \citenamefont {Kapteyn},\ and\ \citenamefont {Christov}}]{Zhou:prl96}%
  \BibitemOpen
  \bibfield  {author} {\bibinfo {author} {\bibfnamefont {J.}~\bibnamefont {Zhou}}, \bibinfo {author} {\bibfnamefont {J.}~\bibnamefont {Peatross}}, \bibinfo {author} {\bibfnamefont {M.~M.}\ \bibnamefont {Murnane}}, \bibinfo {author} {\bibfnamefont {H.~C.}\ \bibnamefont {Kapteyn}}, \ and\ \bibinfo {author} {\bibfnamefont {I.~P.}\ \bibnamefont {Christov}},\ }\href {\doibase 10.1103/PhysRevLett.76.752} {\bibfield  {journal} {\bibinfo  {journal} {Phys. Rev. Lett.}\ }\textbf {\bibinfo {volume} {76}},\ \bibinfo {pages} {752} (\bibinfo {year} {1996})}\BibitemShut {NoStop}%
\bibitem [{\citenamefont {Graml}\ \emph {et~al.}(2023)\citenamefont {Graml}, \citenamefont {Nitsch}, \citenamefont {Seith}, \citenamefont {Evers},\ and\ \citenamefont {Wilhelm}}]{Graml:PRB23}%
  \BibitemOpen
  \bibfield  {author} {\bibinfo {author} {\bibfnamefont {M.}~\bibnamefont {Graml}}, \bibinfo {author} {\bibfnamefont {M.}~\bibnamefont {Nitsch}}, \bibinfo {author} {\bibfnamefont {A.}~\bibnamefont {Seith}}, \bibinfo {author} {\bibfnamefont {F.}~\bibnamefont {Evers}}, \ and\ \bibinfo {author} {\bibfnamefont {J.}~\bibnamefont {Wilhelm}},\ }\href {\doibase 10.1103/PhysRevB.107.054305} {\bibfield  {journal} {\bibinfo  {journal} {Phys. Rev. B}\ }\textbf {\bibinfo {volume} {107}},\ \bibinfo {pages} {054305} (\bibinfo {year} {2023})}\BibitemShut {NoStop}%
\bibitem [{\citenamefont {Haworth}\ \emph {et~al.}(2007)\citenamefont {Haworth}, \citenamefont {Chipperfield}, \citenamefont {Robinson}, \citenamefont {Knight}, \citenamefont {Marangos},\ and\ \citenamefont {Tisch}}]{Haworth:NatPhys07}%
  \BibitemOpen
  \bibfield  {author} {\bibinfo {author} {\bibfnamefont {C.}~\bibnamefont {Haworth}}, \bibinfo {author} {\bibfnamefont {L.}~\bibnamefont {Chipperfield}}, \bibinfo {author} {\bibfnamefont {J.}~\bibnamefont {Robinson}}, \bibinfo {author} {\bibfnamefont {P.}~\bibnamefont {Knight}}, \bibinfo {author} {\bibfnamefont {J.}~\bibnamefont {Marangos}}, \ and\ \bibinfo {author} {\bibfnamefont {J.}~\bibnamefont {Tisch}},\ }\href@noop {} {\bibfield  {journal} {\bibinfo  {journal} {Nat. Phys.}\ }\textbf {\bibinfo {volume} {3}},\ \bibinfo {pages} {52} (\bibinfo {year} {2007})}\BibitemShut {NoStop}%
\bibitem [{\citenamefont {Schmid}\ \emph {et~al.}(2021)\citenamefont {Schmid}, \citenamefont {Weigl}, \citenamefont {Gr{\"o}ssing}, \citenamefont {Junk}, \citenamefont {Gorini}, \citenamefont {Schlauderer}, \citenamefont {Ito}, \citenamefont {Meierhofer}, \citenamefont {Hofmann}, \citenamefont {Afanasiev}, \citenamefont {Crewse}, \citenamefont {Kokh}, \citenamefont {Tereshchenko}, \citenamefont {Güdde}, \citenamefont {Evers}, \citenamefont {Wilhelm}, \citenamefont {Richter}, \citenamefont {Höfer},\ and\ \citenamefont {Huber}}]{Schmid:Nat21}%
  \BibitemOpen
  \bibfield  {author} {\bibinfo {author} {\bibfnamefont {C.~P.}\ \bibnamefont {Schmid}}, \bibinfo {author} {\bibfnamefont {L.}~\bibnamefont {Weigl}}, \bibinfo {author} {\bibfnamefont {P.}~\bibnamefont {Gr{\"o}ssing}}, \bibinfo {author} {\bibfnamefont {V.}~\bibnamefont {Junk}}, \bibinfo {author} {\bibfnamefont {C.}~\bibnamefont {Gorini}}, \bibinfo {author} {\bibfnamefont {S.}~\bibnamefont {Schlauderer}}, \bibinfo {author} {\bibfnamefont {S.}~\bibnamefont {Ito}}, \bibinfo {author} {\bibfnamefont {M.}~\bibnamefont {Meierhofer}}, \bibinfo {author} {\bibfnamefont {N.}~\bibnamefont {Hofmann}}, \bibinfo {author} {\bibfnamefont {D.}~\bibnamefont {Afanasiev}}, \bibinfo {author} {\bibfnamefont {J.}~\bibnamefont {Crewse}}, \bibinfo {author} {\bibfnamefont {K.~A.}\ \bibnamefont {Kokh}}, \bibinfo {author} {\bibfnamefont {O.~E.}\ \bibnamefont {Tereshchenko}}, \bibinfo {author} {\bibfnamefont {J.}~\bibnamefont {Güdde}}, \bibinfo {author} {\bibfnamefont {F.}~\bibnamefont {Evers}}, \bibinfo {author} {\bibfnamefont
  {J.}~\bibnamefont {Wilhelm}}, \bibinfo {author} {\bibfnamefont {K.}~\bibnamefont {Richter}}, \bibinfo {author} {\bibfnamefont {U.}~\bibnamefont {Höfer}}, \ and\ \bibinfo {author} {\bibfnamefont {R.}~\bibnamefont {Huber}},\ }\href@noop {} {\bibfield  {journal} {\bibinfo  {journal} {Nature}\ }\textbf {\bibinfo {volume} {593}},\ \bibinfo {pages} {385} (\bibinfo {year} {2021})}\BibitemShut {NoStop}%
\bibitem [{\citenamefont {Schubert}\ \emph {et~al.}(2014)\citenamefont {Schubert}, \citenamefont {Hohenleutner}, \citenamefont {Langer}, \citenamefont {Urbanek}, \citenamefont {Lange}, \citenamefont {Huttner}, \citenamefont {Golde}, \citenamefont {Meier}, \citenamefont {Kira}, \citenamefont {Koch},\ and\ \citenamefont {Huber}}]{Schubert:NP14}%
  \BibitemOpen
  \bibfield  {author} {\bibinfo {author} {\bibfnamefont {O.}~\bibnamefont {Schubert}}, \bibinfo {author} {\bibfnamefont {M.}~\bibnamefont {Hohenleutner}}, \bibinfo {author} {\bibfnamefont {F.}~\bibnamefont {Langer}}, \bibinfo {author} {\bibfnamefont {B.}~\bibnamefont {Urbanek}}, \bibinfo {author} {\bibfnamefont {C.}~\bibnamefont {Lange}}, \bibinfo {author} {\bibfnamefont {U.}~\bibnamefont {Huttner}}, \bibinfo {author} {\bibfnamefont {D.}~\bibnamefont {Golde}}, \bibinfo {author} {\bibfnamefont {T.}~\bibnamefont {Meier}}, \bibinfo {author} {\bibfnamefont {M.}~\bibnamefont {Kira}}, \bibinfo {author} {\bibfnamefont {S.~W.}\ \bibnamefont {Koch}}, \ and\ \bibinfo {author} {\bibfnamefont {R.}~\bibnamefont {Huber}},\ }\href@noop {} {\bibfield  {journal} {\bibinfo  {journal} {Nat. Photonics}\ }\textbf {\bibinfo {volume} {8}},\ \bibinfo {pages} {119} (\bibinfo {year} {2014})}\BibitemShut {NoStop}%
\bibitem [{\citenamefont {Mandal}\ \emph {et~al.}(2022)\citenamefont {Mandal}, \citenamefont {Rost}, \citenamefont {Pfeifer},\ and\ \citenamefont {Singh}}]{Mandal:OE22}%
  \BibitemOpen
  \bibfield  {author} {\bibinfo {author} {\bibfnamefont {A.}~\bibnamefont {Mandal}}, \bibinfo {author} {\bibfnamefont {J.~M.}\ \bibnamefont {Rost}}, \bibinfo {author} {\bibfnamefont {T.}~\bibnamefont {Pfeifer}}, \ and\ \bibinfo {author} {\bibfnamefont {K.~P.}\ \bibnamefont {Singh}},\ }\href@noop {} {\bibfield  {journal} {\bibinfo  {journal} {Opt. Express}\ }\textbf {\bibinfo {volume} {30}},\ \bibinfo {pages} {45020} (\bibinfo {year} {2022})}\BibitemShut {NoStop}%
\bibitem [{\citenamefont {Shafir}\ \emph {et~al.}(2012)\citenamefont {Shafir}, \citenamefont {Soifer}, \citenamefont {Bruner}, \citenamefont {Dagan}, \citenamefont {Mairesse}, \citenamefont {Patchkovskii}, \citenamefont {Ivanov}, \citenamefont {Smirnova},\ and\ \citenamefont {Dudovich}}]{Shafir:Nat12}%
  \BibitemOpen
  \bibfield  {author} {\bibinfo {author} {\bibfnamefont {D.}~\bibnamefont {Shafir}}, \bibinfo {author} {\bibfnamefont {H.}~\bibnamefont {Soifer}}, \bibinfo {author} {\bibfnamefont {B.~D.}\ \bibnamefont {Bruner}}, \bibinfo {author} {\bibfnamefont {M.}~\bibnamefont {Dagan}}, \bibinfo {author} {\bibfnamefont {Y.}~\bibnamefont {Mairesse}}, \bibinfo {author} {\bibfnamefont {S.}~\bibnamefont {Patchkovskii}}, \bibinfo {author} {\bibfnamefont {M.~Y.}\ \bibnamefont {Ivanov}}, \bibinfo {author} {\bibfnamefont {O.}~\bibnamefont {Smirnova}}, \ and\ \bibinfo {author} {\bibfnamefont {N.}~\bibnamefont {Dudovich}},\ }\href@noop {} {\bibfield  {journal} {\bibinfo  {journal} {Nature}\ }\textbf {\bibinfo {volume} {485}},\ \bibinfo {pages} {343} (\bibinfo {year} {2012})}\BibitemShut {NoStop}%
\bibitem [{\citenamefont {Silaev}\ \emph {et~al.}(2022)\citenamefont {Silaev}, \citenamefont {Romanov},\ and\ \citenamefont {Vvedenskii}}]{Silaev:JPCS22}%
  \BibitemOpen
  \bibfield  {author} {\bibinfo {author} {\bibfnamefont {A.}~\bibnamefont {Silaev}}, \bibinfo {author} {\bibfnamefont {A.}~\bibnamefont {Romanov}}, \ and\ \bibinfo {author} {\bibfnamefont {N.}~\bibnamefont {Vvedenskii}},\ }\href@noop {} {\bibfield  {journal} {\bibinfo  {journal} {Journal of Physics: Conference Series}\ }\textbf {\bibinfo {volume} {2249}},\ \bibinfo {pages} {012004} (\bibinfo {year} {2022})}\BibitemShut {NoStop}%
\bibitem [{sup(2024)}]{suppl}%
  \BibitemOpen
  \href@noop {} {}\bibinfo {howpublished} {See Supplemental Materials for Numerical methods for HHG calculations; Analytical analysis for oriented planar molecule H$_3^{2+}$} (\bibinfo {year} {2024})\BibitemShut {NoStop}%
\bibitem [{\citenamefont {Kong}\ \emph {et~al.}(2022)\citenamefont {Kong}, \citenamefont {Wu}, \citenamefont {Geng},\ and\ \citenamefont {Yu}}]{Kong:FIP22}%
  \BibitemOpen
  \bibfield  {author} {\bibinfo {author} {\bibfnamefont {X.-S.}\ \bibnamefont {Kong}}, \bibinfo {author} {\bibfnamefont {X.-Y.}\ \bibnamefont {Wu}}, \bibinfo {author} {\bibfnamefont {L.}~\bibnamefont {Geng}}, \ and\ \bibinfo {author} {\bibfnamefont {W.-D.}\ \bibnamefont {Yu}},\ }\href {https://www.frontiersin.org/articles/10.3389/fphy.2022.1032671} {\bibfield  {journal} {\bibinfo  {journal} {Front. Phys.}\ }\textbf {\bibinfo {volume} {10}} (\bibinfo {year} {2022})}\BibitemShut {NoStop}%
\bibitem [{\citenamefont {Imasaka}\ \emph {et~al.}(2022)\citenamefont {Imasaka}, \citenamefont {Shinohara}, \citenamefont {Kaji}, \citenamefont {Kaneshima}, \citenamefont {Ishii}, \citenamefont {Itatani}, \citenamefont {Ishikawa},\ and\ \citenamefont {Ashihara}}]{Omasaka:oc22}%
  \BibitemOpen
  \bibfield  {author} {\bibinfo {author} {\bibfnamefont {K.}~\bibnamefont {Imasaka}}, \bibinfo {author} {\bibfnamefont {Y.}~\bibnamefont {Shinohara}}, \bibinfo {author} {\bibfnamefont {T.}~\bibnamefont {Kaji}}, \bibinfo {author} {\bibfnamefont {K.}~\bibnamefont {Kaneshima}}, \bibinfo {author} {\bibfnamefont {N.}~\bibnamefont {Ishii}}, \bibinfo {author} {\bibfnamefont {J.}~\bibnamefont {Itatani}}, \bibinfo {author} {\bibfnamefont {K.~L.}\ \bibnamefont {Ishikawa}}, \ and\ \bibinfo {author} {\bibfnamefont {S.}~\bibnamefont {Ashihara}},\ }\href {\doibase 10.1364/OPTCON.451394} {\bibfield  {journal} {\bibinfo  {journal} {Opt. Continuum}\ }\textbf {\bibinfo {volume} {1}},\ \bibinfo {pages} {1232} (\bibinfo {year} {2022})}\BibitemShut {NoStop}%
\end{thebibliography}%

\end{document}